\documentclass[a4paper,twocolumn,prb]{revtex4-2}

\usepackage[OT1]{fontenc}
\usepackage[latin9]{inputenc}
\usepackage{float}
\usepackage{tikz}
\usepackage{amsmath}
\usepackage{amssymb}
\usepackage{graphicx}
\usepackage{wasysym}
\usepackage[pdfusetitle,
 bookmarks=true,bookmarksnumbered=false,bookmarksopen=false,
 breaklinks=false,pdfborder={0 0 1},backref=false,colorlinks=false]
 {hyperref}

\makeatletter

\pdfpageheight\paperheight
\pdfpagewidth\paperwidth

\providecommand{\tabularnewline}{\\}

\providecolor{lyxadded}{rgb}{0.43,0.88,0.94}
\providecolor{lyxdeleted}{rgb}{1,0,0}
\usetikzlibrary{calc}
\newcommand{\lyxobjectsout}[1]{%
  \bgroup%
  \color{lyxdeleted}%
  \tikz{
    \node[inner sep=0pt,outer sep=0pt](lyxdelobj){#1};
    \draw($(lyxdelobj.south west)+(2em,.5em)$)--($(lyxdelobj.north east)-(2em,.5em)$);
  }
  \egroup%
}

\DeclareRobustCommand{\lyxdisplayobjdeleted}[4][]{%
  \ifx#4\empty\else%
     \texorpdfstring{\leavevmode\\\lyxobjectsout{\parbox{\linewidth}{#4}}}{}%
  \fi%
}
\DeclareRobustCommand{\lyxudisplayobjdeleted}[4][]{%
  \ifx#4\empty\else%
     \texorpdfstring{\leavevmode\\\raisebox{-\belowdisplayshortskip}{%
                \lyxobjectsout{\parbox[b]{\linewidth}{#4}}}}{}%
     \leavevmode\\%
  \fi%
}

\usepackage{bm}
\usepackage{lipsum}
\usepackage{physics}

\usepackage{hyperref}
\hypersetup{
    colorlinks=true,
    linkcolor=blue,
    filecolor=magenta,      
	citecolor=cyan,
    urlcolor=cyan,
    pdftitle={PaperMF_QP}
    }

\makeatother

\begin{document}
\title{Ground state and excitations of quasiperiodic 1D narrow-band moir\'e
systems -- a mean field approach}
\author{Nicolau Sobrosa$^{1}$, Miguel Gon\c{c}alves$^{2}$, Eduardo V. Castro$^{1}$,
Pedro Ribeiro$^{3,4}$, Bruno Amorim$^{1}$}
\affiliation{$^{1}$Centro de Física das Universidades do Minho e do Porto, LaPMET,
Departamento de Física e Astronomia, Faculdade de Ciências, Universidade
do Porto, Rua do Campo Alegre s/n, 4169-007 Porto, Portugal}
\affiliation{$^{2}$Princeton Center for Theoretical Science, Princeton University,
Princeton NJ 08544, USA}
\affiliation{$^{3}$CeFEMA, Instituto Superior Técnico, Universidade de Lisboa,
Av. Rovisco Pais, 1049-001 Lisboa, Portugal}
\affiliation{$^{4}$Beijing Computational Science Research Center, Beijing 100084,
China}
\date{\today}
\begin{abstract}
We demonstrate that a mean field approximation can be confidently
employed in quasiperiodic moir\'e systems to treat interactions and
quasiperiodicity on equal footing. We obtain the mean field phase
diagram for an illustrative one-dimensional moir\'e system that exhibits
narrow bands and a regime with non-interacting multifractal critical
states. By systematically comparing our findings with existing exact
results, we identify the regimes where the mean field approximation
provides an accurate description. Interestingly, in the critical regime,
we obtain a quasifractal charge density wave, consistent with the
exact results. To complement this study, we employ a real-space implementation
of the time-dependent Hartree-Fock, enabling the computation of the
excitation spectrum and response functions at the RPA level. These
findings indicate that a mean field approximation to treat systems
hosting multifractal critical states, as found in two-dimensional
quasiperiodic moir\'e systems, is an appropriate methodology. 
\end{abstract}
\maketitle

\section{Introduction}

With the experimental realization of the twisted bilayer graphene
(tBLG) and other two-dimensional layered materials \citep{wang_correlated_2020,ghiotto_quantum_2021,xu_tunable_2022,li_imaging_2024},
with a fine control over the twist angle \citep{kim_van_2016}, moir\'e
materials have emerged as a new highly tunable platform to study strongly
correlated systems. Moreover, the case of magic angle tBLG is particularly
interesting due to its rich phase diagram, which includes unconventional
superconductivity \citep{cao_unconventional_2018}, correlated insulating
phases \citep{cao_correlated_2018} and topological insulating phases
\citep{nuckolls_strongly_2020}.  The presence of narrow (nearly flat)
bands in the energy spectrum is believed to be a key ingredient in
this plethora of correlated phases. 

Correlated phases on moir\'e materials are usually theoretically studied
through continuum models that assume a plane-wave expansion for the
single-particle states, resulting in an effective periodic structure
\citep{lopes_dos_santos_continuum_2012,bistritzer_moire_2011}. However,
recent research on the narrow band regime of tBLG has surprisingly
suggested that quasiperiodic structures, where the moir\'e wavelength
does not define the unit cell, exhibit critical single-particle states
that delocalize both in real and reciprocal spaces, leading to sub-ballistic
transport properties \citep{goncalves_incommensurability-induced_2021}.
Therefore, the standard continuum model approach must be revised to
include the quasiperiodicity effects.

Quasiperiodicity is not exclusive to tBLG. It appears also in other
two-dimensional moir\'e systems studied recenlty \citep{szabo_mixed_2020,huang_moire_2019,rossignolo_localization_2019}.
Moreover, similar effects are encountered in very distinct types of
systems, including optical lattices \citep{roati_anderson_2008,schreiber_observation_2015,luschen_single-particle_2018,yao_critical_2019,yao_lieb-liniger_2020,gautier_strongly_2021,boers_mobility_2007,modugno_exponential_2009,an_interactions_2021,kohlert_observation_2019},
photonics \citep{lahini_observation_2009,wang_localization_2020,kraus_topological_2012,verbin_observation_2013,verbin_topological_2015,sinelnik_experimental_2020}
and phononics \citep{apigo_observation_2019,ni_observation_2019,cheng_experimental_2020,xia_topological_2020,chen_acoustic_2021,gei_phononic_2020}.
In one dimensional systems, quasiperiodicity induces several novel
properties, such as phase diagrams with mobility edges and a multitude
of localization-delocalization transitions also including critical
states, which are neither localized nor ballistic. In these systems, critical states
delocalize both in real and reciprocal space, having the same nature
as the ones emerging on the narrow-band of the quasiperiodic tBLG.

Establishing the joint role of quasiperiodicity and interactions in
tBLG due to the emergence of sub-ballistic states is yet to be accomplished.
Such a study requires highly-scalabe numerical methods able to take
electron-electron interations into account in quasiperiodic systems.
In two-dimensions, the lack of such numerical methods turns this problem
almost prohibitive. In tBLG, real space approaches usually rely on
the tight-binding approach \citep{trambly_de_laissardiere_localization_2010}
and adopt the Hartree-Fock approximation to treat the correlated phases
\citep{sboychakov_charge_2022,breio_chern_2023,faulstich_interacting_2023,gonzalez_magnetic_2021,gonzalez_time-reversal_2020,vahedi_magnetism_2021}.
However, on the few studies that use this methodology, the structures
considered are always commensurate, \emph{i.e.} periodic, far from
the quasiperiodic regime and the emergence of the non-interacting
sub-ballistic states.

For one-dimensional (1D) systems, however, the Density Matrix Renormalization
Group (DMRG) \citep{schollwock_density-matrix_2005} is a numerically
exact method capable of obtaining the exact ground-state of large-scale
interacting systems. In Ref.~\citep{goncalves_incommensurability_2023},
the authors performed an in-depth study of the interacting phase diagram
of a 1D system which exhibits a localization transition between extended
states and multifractal critical states due to quasiperiodic modulation
that also creates a narrow band at the Fermi level. Surprisingly,
those critical states are unstable, when adding electron-electron
interactions to a quasi-fractal charge order, characterized by an
extremely large number of wave-vectors, diverging for an infinitesimal
interaction. This phase does not exist for periodic structures nor
for the extended regime, being a direct consequence of the presence
of the quasiperiodicity. This is one of the first evidences that new
correlated phases may be stabilized due to the sole effect of quasiperiodicity,
enhancing our belief that the quasiperiodic nature of tBLG must be
taken into account. However, such an approach is only feasible for
1D systems. 

\begin{figure}
\centering{}\includegraphics[width=0.9\columnwidth]{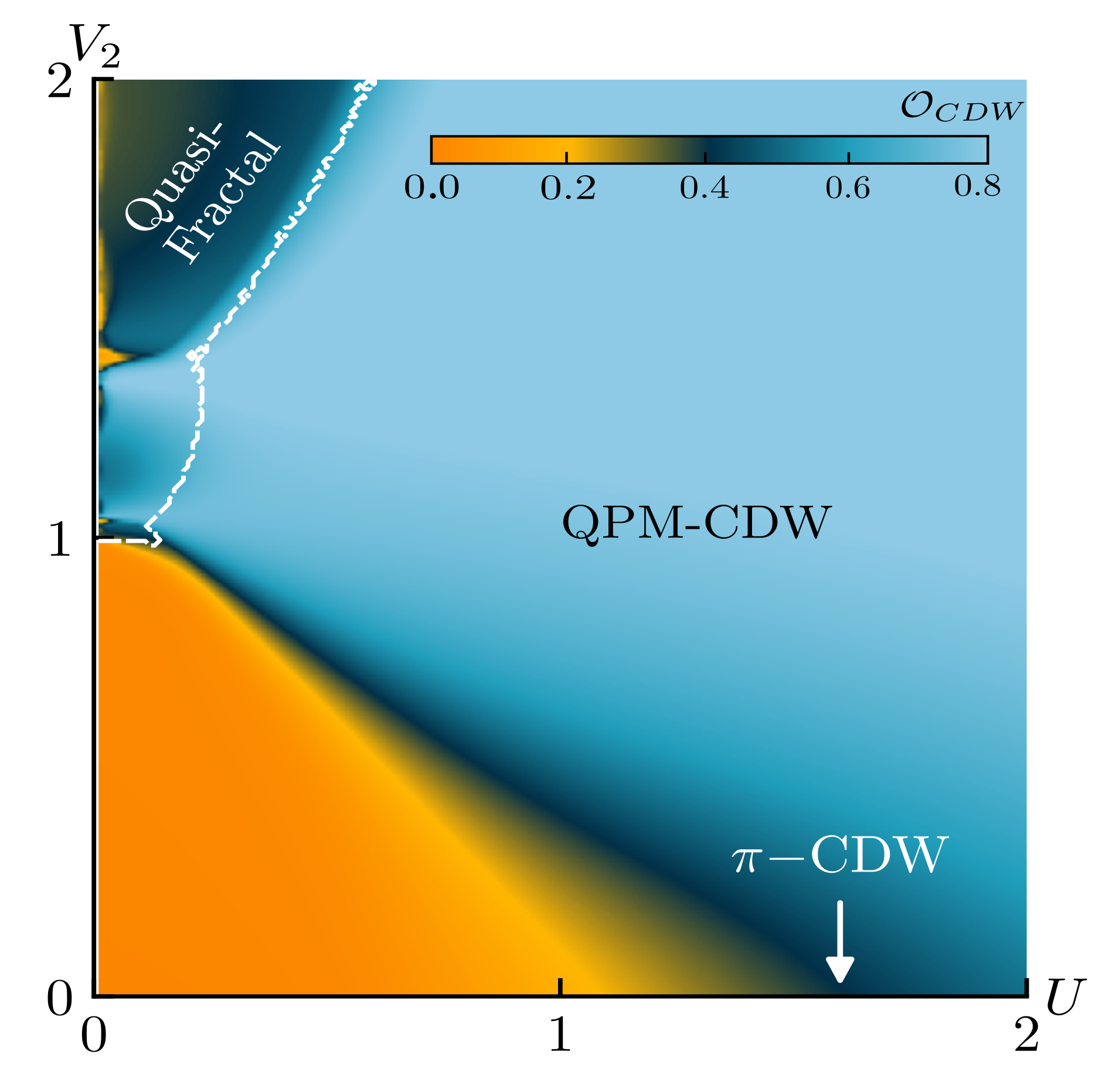}\caption{\label{fig:PD_QP}Mean field phase diagram for the quasiperiodic case.
 The white line marks the points where the Fourier transform of the
charge density fluctuations change from extend to localized behavior.
Each color maps the magnitude of the  order parameter, $\mathcal{O}_{CDW}=\max\boldsymbol{n}-\min\boldsymbol{n},$
where $\boldsymbol{n}$ is the vector of the charge density. $\pi$-CDW
corresponds to the Charge Density Wave with order only at $k=\pi.$
QPM-CDW to the Quasiperiodic moir\'e charge density wave, where a finite
number of wave vectors are present. Quasi-fractal corresponds to the
regime of the charge density wave where an extremely large number
of wave-vectors are present in the fluctuations. A system with size
$N=504$ and modulation period $\tau=\frac{293}{504}$ was used.}
\end{figure}

The goal of this paper is to assert whether a real-space mean field
approach is a valid method to treat correlated phases, when considering
systems with the particularities of quasiperiodic tBLG, namely the
narrow bands and critical states. To check the validity of such methodology,
we used the 1D model of Ref.~\citep{liu_localization_2015}, that
exhibits those two features. By comparing with DMRG exact results
\citep{goncalves_incommensurability_2023}, we  establish the effectiveness
of the mean field treatment:  for the critical regime, we found an
astonishing resemblance with the exact results, with the mean field
approximation being able to obtain the quasi-fractal  charge density
wave (CDW) phase for any infinitesimal interaction, as shown in Fig.~\ref{fig:PD_QP}.
With the aid of a real-space implementation of the time-dependent
Hartree-Fock (tdHF) method, equivalent to the Random Phase Approximation
(RPA) \citep{co_introducing_2023}, we were able to study the excitation
spectrum of the system, unobtainable from the point of view of DMRG,
as well as different generalized susceptibilities, in the frequency-momentum
domain. We arrive at the so-called Bethe Salpether equation for the
4-point correlation functions and to an effective two-electron Hamiltonian,
in the particle-hole sector of excitations. The study of the excitations
gives us information about the collective modes of the system. 

The paper is structured as follows: In Sec.~\ref{sec:Model-and-methods}
we introduce the Hamiltonian of the system as well as the single-particle
properties that arise from the quasiperiodic modulation. Also, we
introduce the mean field approximation and the tdHF approach. We describe
briefly the observables that we used to describe the ordered phases.
We proceed to present the mean field results in Sec.~\ref{sec:Mean-Field-Results}
with an in-depth description of the several phases that we have found.
In Sec.~\ref{sec:Beyond-mean-field} we present the excitation spectrum
for the extended and critical regimes as well as some two-particle
wavefunctions. We also show the charge response functions with, and
without, interactions to explain some of the instabilities of the
system. In Sec.~\ref{sec:Conclusions} the key results are summarized
and some conclusions are drawn. We also include two appendices: in
Appendix~\ref{sec:Time-dependent-Hartree-Fock} we give a detailed
derivation of the tdHF from a linear response perspective; in Appendix~\ref{sec:Hubbardf}
we cross check the method by studying the excitation spectrum of the
1D Hubbard model and the spin-wave spectrum using the tdHF method;
in Appendix~\ref{sec:ICDW} we compute the excitation spectrum of
the system in the clean limit away from half-filling, where we obtain
the \emph{phason} mode.
\begin{figure}
\centering{}\includegraphics[width=0.98\columnwidth]{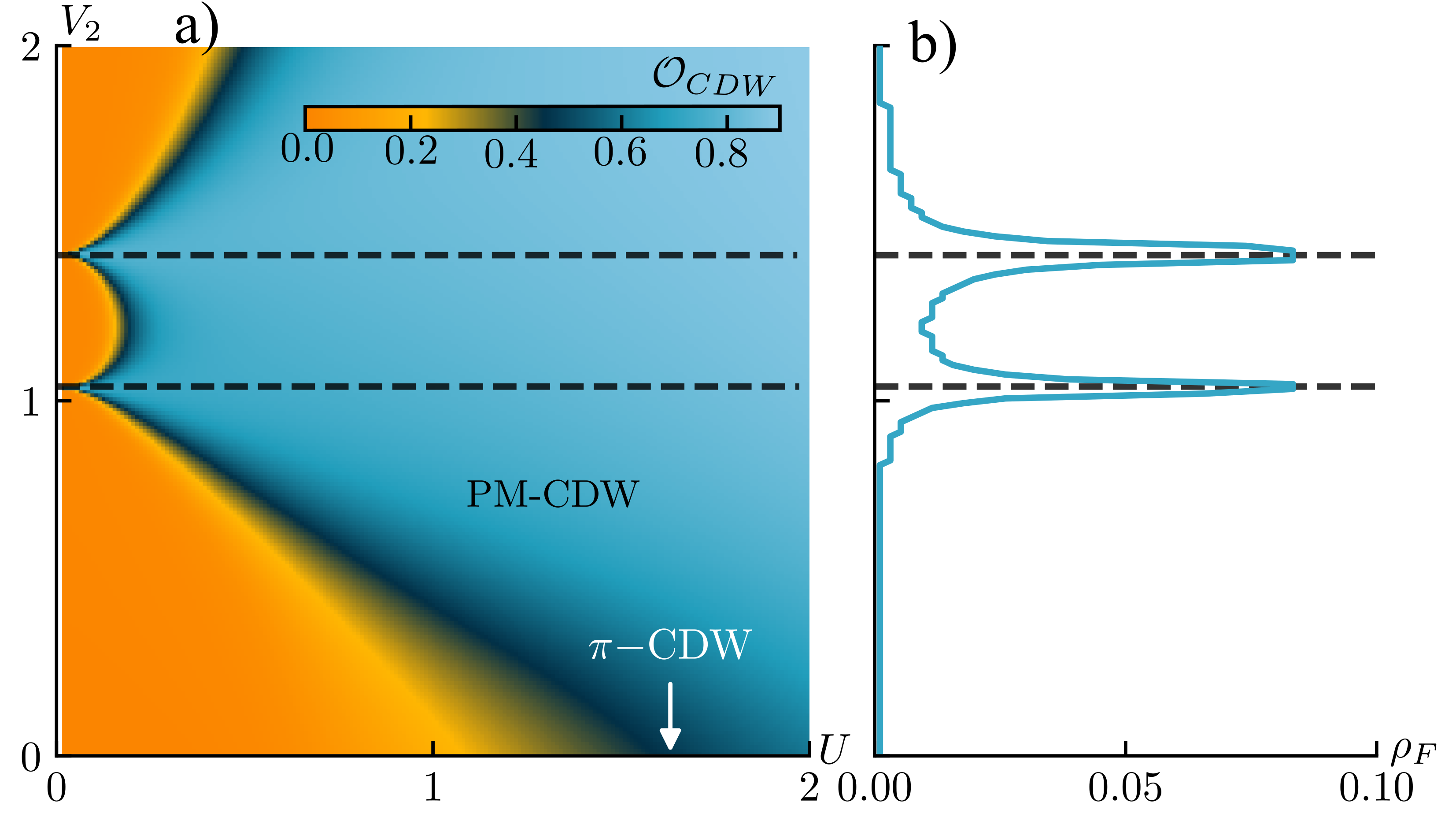}\caption{\label{fig:PD_P}a) Mean field phase diagram for the periodic case. Each
color maps the magnitude of the order parameter, $\mathcal{O}_{CDW}$.
The horizontal dashed lines mark the value of the modulation strength
where the bands become narrower, with a higher density of states,
thus reducing enhancing the order paramter. $\pi$-CDW corresponds
to the Charge Density Wave with order only at $k=\pi.$ PM-CDW to
the Periodic moir\'e Charge Density Wave, where a finite number of  possible
wavevectors (in this case 12) are present in the charge density fluctuations.
b) Density of states of the non-interacting Hamiltonian, at the Fermi
level as function of the periodic modulation strength, $V_{2}$. A
system size $N=480$ and modulation period $\tau=\frac{7}{12}$ was
used. }
\end{figure}

\section{Model and methods\label{sec:Model-and-methods}}

We consider a 1D tight-binding chain of spinless fermions with a quasiperiodic
spatial modulation of the hoppings \citep{liu_localization_2015}.
We also consider a nearest neighbors repulsive interaction with magnitude
$U>0$. The complete Hamiltonian reads
\begin{equation}
H=\sum_{j}t_{j}c_{j}^{\dagger}c_{j+1}+\text{h.c.}+U\sum_{j}n_{j}n_{j+1},\label{eq:H_int}
\end{equation}
where $c_{j}^{\dagger}$ creates an electron at site $j$, $t_{j}=-t-V_{2}\cos\left(2\pi\tau\left(j+\frac{1}{2}\right)+\phi\right)$
is the hopping amplitude from site $j+1$ to $j$, which has a quasiperiodic
modulation strength $V_{2}$, period $\tau^{-1}$ and phase shift
$\phi$. Throughout the work, we consider $t=1$ and all energy scales
are measured through the hopping strength. For irrational $\tau$,
the periodicity of the hopping term becomes infinite. We consider
the model at half-filling and periodic boundary conditions (PBC).
To perform a finite size scaling analysis, we used a well established
procedure, approximating the irrational $\tau$ by a rational number,
$\tau_{p,N}=p/N,$ where $N$ is the size of the considered finite
chain, that contains one unit cell, and $p$ is a co-prime number
of $N$. We then consider a sequence of approximants, $\tau_{p,N}$,
of increasing system size $N$. We consider $\tau=\frac{1}{2}+\delta$,
with $\delta\approx0.0812$, which ensures the formation of a narrow
band exactly at the center of the energy spectrum. In this way, the
moir\'e pattern exhibits a moir\'e length, $L_{M}=\left(\tau-\frac{1}{2}\right)^{-1}=\delta^{-1}$,
of approximately $12$ atoms. Furthermore, there is a localization
transition at $V_{2}=1$ where all states go from extended to multifractal
critical states on increasing $V_{2}$. The latter are characterized
by being delocalized in reciprocal space as well as in real space.

Without any quasiperiodic modulation, the exact interacting ground
state exhibits a CDW with order at wavevector $k=\pi$, for a sufficiently
large $U$, forming a unit cell composed by two atoms. To ensure that
no boundary defects arise in the mean field calculations, we consider
systems with an even number of sites. To maximize the available system
sizes, we do not consider an exact sequence of approximants, having
in mind that small variations on $\tau$ should not change the physical
properties of the system. The series of the chosen approximants is
given in Tab.~\ref{tab:approx}. We compare the quasiperiodic systems (QPS) with periodic ones (PS)
by considering $\tau\equiv\tau_{c}=\frac{7}{12},$ where the system's
unit cell, when repeated across the chain, coincides with the moir\'e
pattern wavelength of $12$ sites. 
\begin{table}[H]
\begin{centering}
\begin{tabular}{|c|c|c|c|c|c|c|c|c|}
\hline 
$N$ & 112 & 196 & 308 & 504 & 1008 & 1476 & 2008 & 2500\tabularnewline
\hline 
\hline 
$\tau_{p,N}$ & $\frac{65}{112}$ & $\frac{113}{196}$ & $\frac{179}{308}$ & $\frac{293}{504}$ & $\frac{587}{1008}$ & $\frac{859}{1478}$ & $\frac{1167}{2008}$ & $\frac{1453}{2500}$\tabularnewline
\hline 
\end{tabular}
\par\end{centering}
\caption{\label{tab:approx}List of chosen approximants to the irrational $\tau\approx$0.5812.}
\end{table}

The results were obtained using a variational mean field-decoupling
of the interacting Hamiltonian. We propose a mean field (non-interacting)
Hamiltonian, 
\begin{equation}
H_{MF}=H_{0}+\sum_{ij}\epsilon_{ij}c_{i}^{\dagger}c_{j},\label{eq:HMF_general}
\end{equation}
with $H_{0}$ the non-interacting part of the full Hamiltonian in
Eq.~\eqref{eq:H_int}, and where $\epsilon_{ij}$ are generic mean-field
variational parameters that couple the sites $j$ and $i$- Then we
minimize the internal energy of the full Hamiltonian with respect
to the ground state of the mean field Hamiltonian, $\left\langle H\right\rangle _{MF}$,
using the Gibbs-Bogoliubov-Feynamn inequality as the variational principle
\citep{kuzemsky_variational_2015}. For the considered interaction,
the mean field Hamiltonian takes the form
\begin{equation}
H_{MF}=H_{0}+\sum_{i}\epsilon_{i}c_{i}^{\dagger}c_{i}+\sum_{i}\Delta_{i}c_{i}^{\dagger}c_{i+1}+h.c.,\label{eq:HMF_specific}
\end{equation}
with 
\begin{align}
\epsilon_{i} & =U\left(\left\langle n_{i-1}\right\rangle _{MF}+\left\langle n_{i+1}\right\rangle _{MF}\right)\nonumber \\
\Delta_{i} & =-U\left\langle c_{i+1}^{\dagger}c_{i}\right\rangle _{MF},\label{eq:MF}
\end{align}
where $\epsilon_{i}$'s correspond to the Hartree term and $\Delta_{i}$'s
are the Fock ones. Note that $\left\langle \dots\right\rangle _{MF}$
represents the average value of an operator in the ground state of
the mean field Hamiltonian. Therefore, Eq.~\eqref{eq:MF} represents
a set of self-consistent equations that must be solved iteratively
using the eigenstates of Eq.~\eqref{eq:HMF_specific}. Specifically,
if $\mathcal{\mathcal{U}}_{i,\alpha}$ is the $i$-th component of
the $\alpha$-th eigenstate, we may write the following mean field
expectation values,
\begin{align}
\left\langle n_{i}\right\rangle _{MF} & =\sum_{\alpha\text{occ}}\left|\mathcal{\mathcal{U}}_{i,\alpha}\right|^{2}\nonumber \\
\left\langle c_{i+1}^{\dagger}c_{i}\right\rangle _{MF} & =\sum_{\alpha\,\text{occ}}\mathcal{U}_{i+1,\alpha}^{*}\mathcal{U}_{i,\alpha},\label{eq:MF_vecs}
\end{align}
where we have  considered the zero temperature limit. We solve self-consistently
Eq.~\eqref{eq:MF_vecs} until the internal energy and the charge
density achieve the convergence criteria, set through the absolute
error, which must be below $10^{-8}$. 

To study the excitation spectrum of the system and fluctuations of
the mean field solution, we employ a real-space implementation of
the time-dependent Hartree Fock, based on the linear response of the
reduced density matrix \citep{bernasconi_optical_2012,dreuw_single-reference_2005}.
We start by considering the interacting Hamiltonian in the basis that
diagonalizes the mean-field Hamiltonian,
\begin{equation}
H=\sum_{a}E_{a}^{(\text{MF})}d_{a}^{\dagger}d_{a}+\sum_{abcd}U_{cd}^{ab}d_{a}^{\dagger}d_{b}^{\dagger}d_{c}d_{d},\label{eq:HMF_diag}
\end{equation}
where $d_{a}^{\dagger}$ creates a state with energy $E_{a}^{(\text{MF})}$
and $U_{cd}^{ab}$ are the interaction matrix elements, with $a,b,c,d$
indices of the single-particle mean-field states. We add a time-dependent
perturbation of the form 
\[
A_{p}(t)=F(t)\sum_{ab}A_{ab}d_{a}^{\dagger}d_{b},
\]
where $A_{ab}$ are the perturbation matrix elements and $F(t)$ encompasses
the full time dependence. Then, we apply the linear response approximation
to the time-dependent reduced density matrix, 
\begin{equation}
\rho_{ba}\left(t\right)=\left\langle d_{a}^{\dagger}\left(t\right)d_{b}\left(t\right)\right\rangle .\label{eq:rdm}
\end{equation}
At zero temperature, restricted to the electron-hole sector, the correction
to first order in the applied perturbation, may be written as 
\begin{equation}
\boldsymbol{\rho}^{(1)}\left(\omega\right)=\left(\hbar\omega\boldsymbol{I}-\boldsymbol{H}_{eh}\right)^{-1}\boldsymbol{V}_{p}\left(\omega\right),\label{eq:rdmCorrection}
\end{equation}
where we have taken a Fourier transform in the time coordinate, with
$H_{eh}$ is the electron-hole Hamiltonian, 
\begin{equation}
\boldsymbol{H}_{eh}=\left(\begin{array}{cc}
\boldsymbol{R} & \boldsymbol{C}\\
-\boldsymbol{C}^{*} & \boldsymbol{-R}^{*}
\end{array}\right),\label{eq:H_eh}
\end{equation}
with elements
\begin{align}
\boldsymbol{R}_{e^{\prime}o^{\prime}}^{eo} & =\left(E_{o}-E_{e}\right)\delta_{ee^{\prime}}\delta_{oo^{\prime}}+\left(U_{eo^{\prime}}^{e^{\prime}o}+U_{o^{\prime}e}^{oe^{\prime}}-U_{o^{\prime}e}^{e^{\prime}o}-U_{eo^{\prime}}^{oe^{\prime}}\right)\label{eq:Elements}\\
\boldsymbol{C}_{o^{\prime}e^{\prime}}^{eo} & =U_{ee^{\prime}}^{oo^{\prime}}+U_{e^{\prime}e}^{o^{\prime}o}-U_{ee^{\prime}}^{o^{\prime}o}-U_{e^{\prime}e}^{oo^{\prime}},\nonumber 
\end{align}
where $ee^{\prime}$ are indices for empty states and $oo^{\prime}$
correspond to occupied states. The eigenvalues of this electron-hole
Hamiltonian correspond to the excitation spectrum, while the eigenvectors
correspond to the different instability channels of the system. Also,
this two-particle Hamiltonian enables the computation of the generalized
susceptibility between two operators, $\boldsymbol{A}$ and $\boldsymbol{B}$,
as given by 
\begin{equation}
\chi_{BA}\left(\omega\right)=\left[\boldsymbol{B}_{eo}\,\,\,\boldsymbol{B}_{oe}\right]\left[\begin{array}{cc}
\hbar\omega\boldsymbol{I}-\boldsymbol{R} & -\boldsymbol{C}\\
\boldsymbol{C}^{*} & \hbar\omega\boldsymbol{I}+\boldsymbol{R}^{*}
\end{array}\right]^{-1}\left[\begin{array}{c}
\boldsymbol{A}_{eo}\\
-\boldsymbol{A}_{oe}
\end{array}\right].\label{eq:gen_Sus}
\end{equation}

To fully characterize the ordered phases we used different observables.
First, we compute the system gap by considering the energy difference
between the first empty state and the last occupied state. To characterize
the CDW, we compute the fluctuations around the average value, $\left\langle \delta n_{m}\right\rangle =\left\langle n_{m}\right\rangle -\frac{1}{2}$,
and also its Fourier transform
\begin{equation}
\left\langle \delta n_{k}\right\rangle =\frac{1}{\sqrt{N}}\sum_{m}e^{ikm}\left\langle \delta n_{m}\right\rangle ,\label{eq:flutuacoes}
\end{equation}
that signals the formation of an ordered phase at wavevector $k$.
Regarding the order parameter, we have defined it to be
\begin{equation}
\mathcal{O}_{CDW}=\max_{i}\left\langle n_{i}\right\rangle -\min\left\langle n_{i}\right\rangle .\label{eq:OCDW}
\end{equation}
We chose this definition due to the peculiar structure of the CDWs,
where in certain regions of the phase diagram, the modulation is very
localized and a more common approach such as the average over the
difference of charge in neighboring sites dilutes this type of structure.
To complement this quantity, we also computed a generalization of
the Inverse Participation Ratio, defined as 
\begin{equation}
\text{IPR}\left(\left\langle \delta\boldsymbol{n}\right\rangle \right)=\left(\sum_{k}\left|\left\langle \delta n_{k}\right\rangle \right|^{4}\right)/\left(\sum_{k}\left|\left\langle \delta n_{k}\right\rangle \right|^{2}\right)^{2}.\label{eq:IPR}
\end{equation}
This quantity allows the distinction between extended and localized
density fluctuations in momentum-space. If only a small (and intensive)
number of $k-$values contribute to the fluctuations, the IPR should
scale with the chain size as $N^{0}$ while for extended states it
scales as $N^{-1}$ \citep{goncalves_incommensurability_2023}.

\section{Mean Field Results\label{sec:Mean-Field-Results}}

In the $V_{2}=0$ limit, the system exhibits a CDW with order at the
wavector $k=\pi$, corresponding to the $\pi-$CDW phase, for any
interaction strength $U>0$, with a finite gap for all interaction
strength. This result contrasts with the exact result where a phase
transition between a (gapless) Luttinger liquid and a $\pi-$CDW,
with a finite gap, occurs for a finite critical interaction, $U_{c}=2$
\citep{cazalilla_one_2011}. Using a reciprocal-space implementation
of our method we have checked that the gap is exponentially suppressed
for low values of the interaction strength, $\Delta\sim\exp\left(-\frac{1}{U}\right)$,
which is in perfect agreement with mean field analytical calculations
\citep{shankar_renormalization-group_1994}. In this regime, the mean
field method is not accurate, although we can describe well the structure
of the CDW, despite missing the phase transition.

\subsection{Extended States upon interaction}

\begin{figure}
\begin{centering}
\includegraphics[width=0.98\columnwidth]{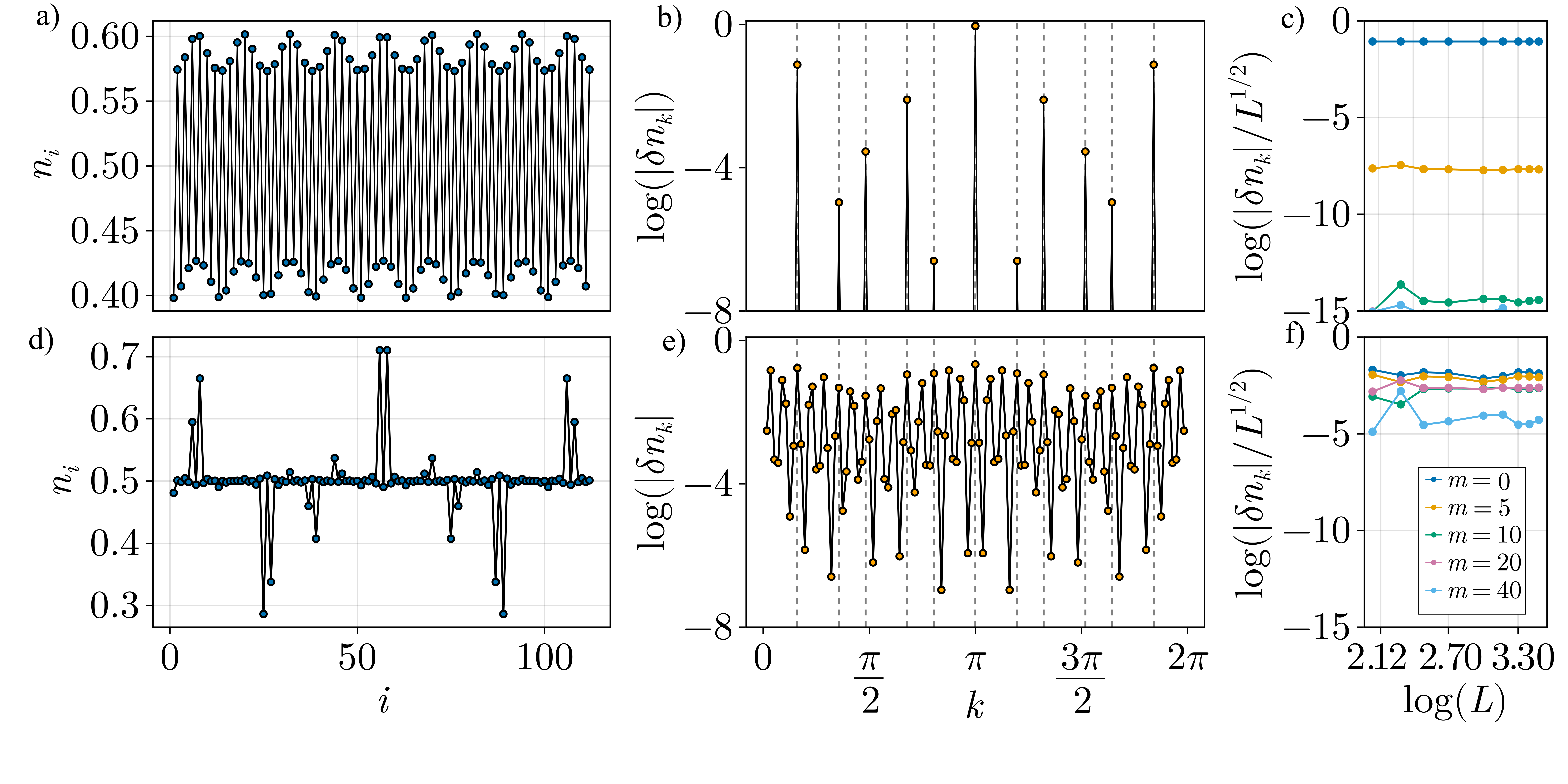}
\par\end{centering}
\caption{\label{fig:MF_QP}Mean field results of the CDW phase for a system
size of $N=112$, for different quasiperiodic modulation strengths.
a)~Real space charge density modulation for $V_{2}=0.1$ and $U=1$.
b)~Fourier transform of the fluctuations of the charge density for
$V_{2}=0.1$ and $U=1$. The vertical dashed lines correspond to the
position of the peaks, $K_{n}=\pi+2\pi\tau n$. d)Real space charge
density modulation for $V_{2}=2.0$ and $U=0.1$. e) Fourier transform
of the fluctuations of the charge density for $V_{2}=2.0$ and $U=0.1$.
Vertical dashed lines correspond to the position of the peaks, $K_{n}=\pi+2\pi\tau n$.
Panels c) and f) correspond to the finite size scaling analysis for
selected peaks of panels b) and e), respectively. The selected peaks
correspond to $K_{m}=\pi+2\pi\tau m,$ with $m=0,5,10,20,40$.}
\end{figure}

In the extended regime, $V_{2}<1$, any interaction strength, $U>0$,
generates a CDW with an exponentially suppressed gap for small $U$.
The charge density exhibits a more complex structure, where a finite
number of wavevectors are present in the Fourier transform of the
charge density fluctuations. In Fig.~\ref{fig:MF_QP}a), we represent
an example of the charge density distribution for $V_{2}=0.1$ and
$U=1$, for a chain with $112$ atoms. In Fig.~\ref{fig:MF_QP}b),
we show the corresponding Fourier transform of the fluctuations where
only a finite number of wavevectors contribute and are described by
\[
K_{n}=\pi+2\pi\tau n,
\]
indicated by the dashed lines. In this regime, the number of peaks
is always finite and saturates with increasing system size. In Fig.
\ref{fig:MF_QP}c) we show the value of $\log(|\delta n_{k}|)$ as
a function of system size for selected peaks, $n\in[0,5,10,20,40],$where
the first two are, clearly, converged, while the latter three are
below machine precision, implying that they do not contribute to the
CDW. We call this phase a quasiperiodic-moir\'e CDW, since the charge
modulation has the same moir\'e pattern as the modulation of the hoppings.

\begin{figure}
\centering{}\includegraphics[width=1\columnwidth]{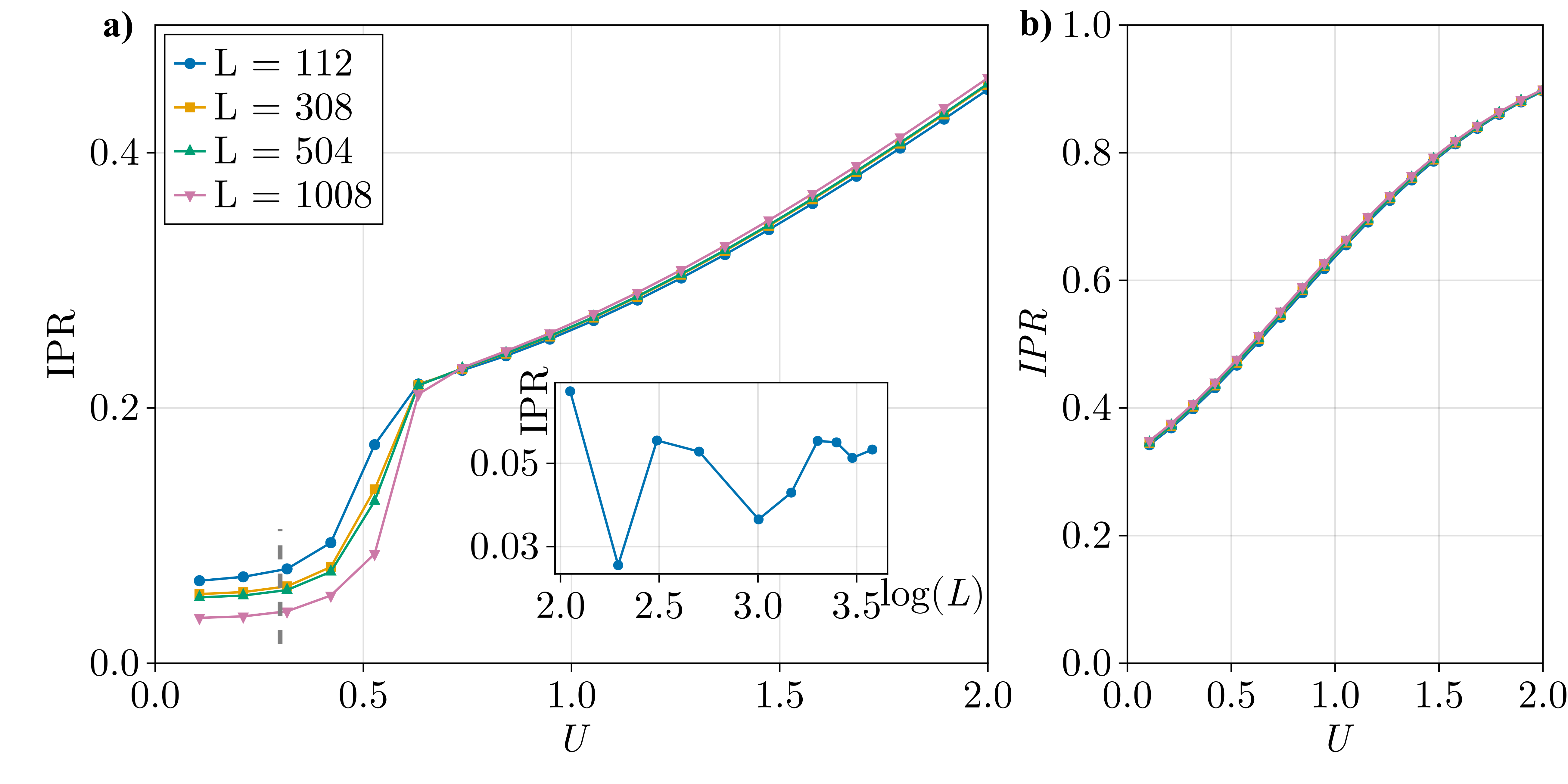}\caption{\label{fig:IPRV2=00003D0.5}IPR of the charge density fluctuations
as a function of the interaction strength for quasiperiodic modulation
a) $V_{2}=0.5$ and b) $V_{2}=2.0$, for different chain sizes. The
inset shows a detailed finite size scaling analysis for a selected
value of the interaction, marked as a dashed line in panel a). }
\end{figure}

To further study the fluctuations of the charge density, we computed
the IPR as a function of the interaction strength, for a fixed $V_{2}$
in the extended region. The result is shown in Fig.~\ref{fig:IPRV2=00003D0.5}
b) for different system sizes. For every value of $U$, the value
of the IPR is converged with the system size,indicating that the fluctuations
are always localized, \emph{i.e}, only a small number of wavevectors
(that does not scale with the system size) contribute to the charge
order. The increase in the IPR value with increasing interaction strength
indicates that the number of wavevectors is being suppressed and the
order is being dominated by the $k=\pi$ wavevector as it is expected
in the $U\to\infty$ (classical) limit.

\subsection{Critical states with interaction}

In the critical regime, $V_{2}>1$, we found an impressively satisfactory
result, when compared to the exact ones. For an infinitesimal interaction
strength, the system is unstable, at the mean field level, to a gapped
CDW with finite (non-exponentially suppressed) order parameter. Furthermore,
the structure of the CDW is precisely the same as in the exact solution
found in Ref.~\citep{goncalves_incommensurability_2023}. The fluctuations
are very localized and spaced apart from one another, as can be seen
in Fig.~\ref{fig:MF_QP}d) for $V_{2}=2.0$ and $U=0.1$. That result
is also corroborated with the Fourier transform of the fluctuations,
Fig.~\ref{fig:MF_QP}e), where we obtain a finite contribution from
a very large number of wavevectors. As can be seen in Fig.~\ref{fig:MF_QP}f),
all shown peaks are well converged with the system size and above
machine precision. This phase corresponds to the quasiperiodicity-induced
quasifractal CDW that is found in the DMRG result of Ref.~\citep{goncalves_incommensurability_2023}.

In Fig.~\ref{fig:IPRV2=00003D0.5}a), we show how the IPR evolves
with the system size as a function of the interaction strength,  for
$V_{2}=2$. There are clearly two regimes. For $U>U^{*}$, the CDW
has the same structure as in the extended regime, with an IPR converged
even for the smallest system sizes, indicating the contribution of
only a small number of wavevectors. However, for $U<U^{*}$, the IPR
decreases abruptly, indicating that a much larger number of wavevectors
are now contributing. As can be seen in the inset of Fig.~\ref{fig:IPRV2=00003D0.5}a),
even though the number of wavevectors can be large, their number is
finite since the IPR converges for larger system sizes to a nonzero
value. As $U$ decreases, the number of wavectors becomes larger and
larger, as signaled by the reduction of the IPR. This is the quasifractal
CDW phase mentioned above, and $U^{*}$ can be identified with the
crossover to the QPM-CDW found in Ref.~\citep{goncalves_incommensurability_2023}.
The crossover interaction $U^{*}$ is shown as a dashed, white line
in Fig.~\ref{fig:PD_QP}. The present mean-field results also corroborate
the conjecture, put forward in Ref.~\citep{goncalves_incommensurability_2023},
that $\text{IPR}\rightarrow0$ when $U\rightarrow0$.

\subsection{Periodic structures}

Considering now $\tau_{c}=7/12$, the system becomes periodic and
the Bloch's theorem's applies for any hopping modulation strength.
On this scenario, we obtain a CDW with a modulation only inside each
unit cell . The exponentially suppressed gap (and order parameter),
found in the clean limit, extends to every value of $V_{2}$, as may
be seen in Fig.~\ref{fig:PD_P}. Since the density of states changes
rapidly with $V_{2}$, due to the flattening of the bands, the response
to the interaction is different as a function of $V_{2}$. The clean
limit result, $\Delta(U)\sim\exp\left(-\frac{U_{\text{typ}}}{U}\right),$
with $U_{\text{typ}}^{-1}=2\rho_{F}$, and $\rho_{F}$ the density
of states at the Fermi level, is valid for every $V_{2}$, with the
apparent modulation simply being a result of the flattening of the
bands, and an increase in the density of states. The exact result
found in Ref.~\citep{goncalves_incommensurability_2023} shows a
phase transition that exhibits the same relation as $U_{\text{typ}}^{-1}$.
However, similar to the clean limit, the mean field method is not
able to correctly obtain the exact phase transition. Nonetheless,
with respect to the charge distribution, the two methods agree well,
capturing correctly the structure of the CDW.

\begin{figure}
\begin{centering}
\includegraphics[width=0.99\columnwidth]{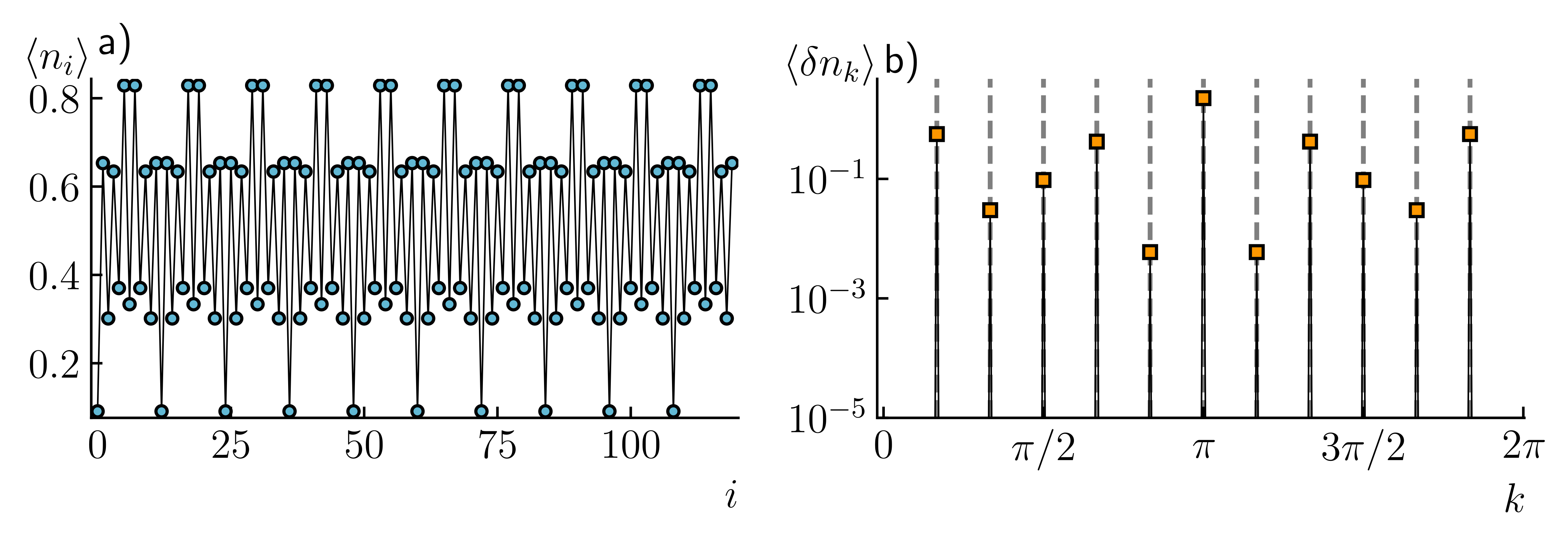}
\par\end{centering}
\caption{\label{fig:PerChaDis}a) Charge distribution for a periodic system,
with $\tau_{c}=\frac{7}{12},$for $U=1$ and $V_{2}=1$, where the
charge distribution has a periodicity of $12$ atoms. b) Fourier transform
of the charge fluctuations. The dashed lines correspond to $K_{n}=\pi+2\pi\tau_{c}n$
with $n=-5,\cdots,6$.}

\end{figure}
In Fig. \ref{fig:PerChaDis}a), we show one example of the charge
distribution for the periodic regime, with $\tau_{c}=\frac{7}{12}$,
$V_{2}=1.0$ and $U=2.0$ for a chain of size $L=120$. In this regime,
the charge distribution has the same structure for every pair $(U,V_{2})$..
In Fig.~\ref{fig:PerChaDis}b) we show the corresponding Fourier
transform of the charge density fluctuations. The series of peaks,
marked as dashed lines, are given by 
\[
K_{n}=\pi+2\pi\tau_{c}n.
\]
Differently from the quasiperiodic case, the set of $K_{n}$ only
consists on $12$ different values, for any hopping modulation strength.
Therefore, the CDW has always the same period of the moir\'e pattern
and, in particular, the same size as the unit cell, matching correctly
with the exact solution of Ref.~\citep{goncalves_incommensurability_2023}.

\section{Beyond mean field \label{sec:Beyond-mean-field}}

With the mean field ground state well established, we proceed to the
study of the excitation spectrum. Through the tdHF theory (see Appendix~\ref{sec:Time-dependent-Hartree-Fock}),
we obtain the eigenvalues of the two-particle Hamiltonian of Eq.~\eqref{eq:H_eh}.
Since the mean field ground state has a charge modulation, we focus
our analysis into the charge response function. Then, analyzing the
excitation spectrum, we are able to identify the collective modes
of the system. In Appendix~\ref{sec:Hubbardf}, we study the excitation
spectrum of the textbook Hubbard model in 1D, to crosscheck the analysis
and exemplify how it is performed. The spin-wave spectrum is obtained,
providing clear evidence that tdHF method is, indeed, capable of obtaining
such modes.

\subsection{Excitation spectrum}

\begin{figure}
\begin{centering}
\includegraphics[width=0.98\columnwidth]{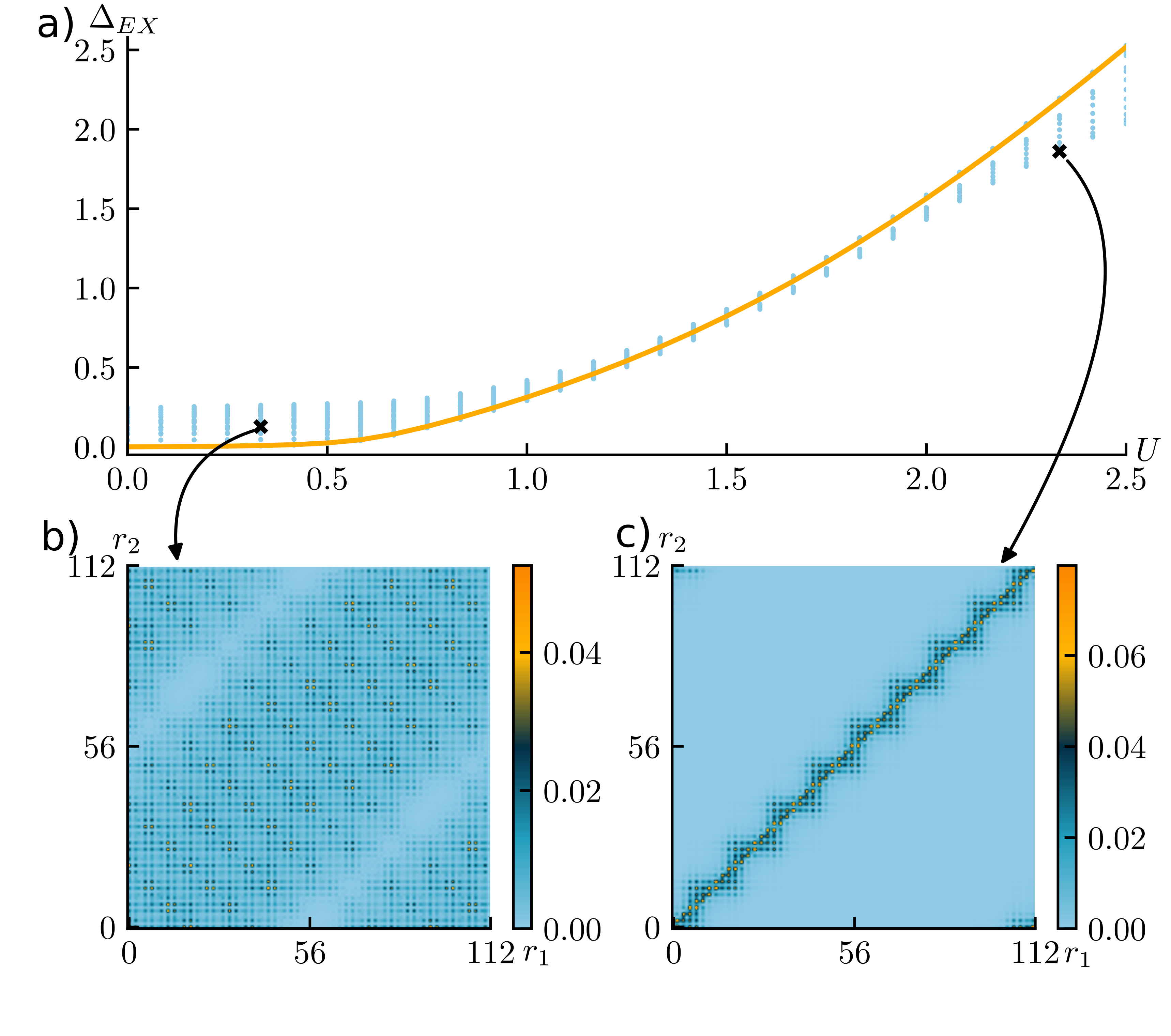}
\par\end{centering}
\centering{}\caption{\label{fig:ExSp_V2=00003D0.5}Excitation spectrum for the extended
regime, $V_{2}=0.5$ and system size $N=112$. a)~Blue dots mark
the low-energy excitation energies and the orange line corresponds
to the mean-field gap. Real-space representation of an eigenvector
of an excitation with energy above the gap in b) and below the gap
in c).}
\end{figure}

For $U=0$, the excitation spectrum is described through the difference
between electron and hole energies, $\Delta_{EX}=\epsilon_{h}-\epsilon_{e}$.
Therefore, the minimal excitation energy is the mean field gap. When
introducing interactions, excitations whose energy is below the gap
appear, which correspond to collective modes of the system. We restrict
ourselves only to the low-energy excitations that correspond to particle-hole
excitations inside the narrow band at the Fermi level, in the non-interacting
limit. In Fig.~\ref{fig:ExSp_V2=00003D0.5}a), we show the excitation
spectrum as a function of the interaction strength, where the orange
line marks the mean field gap. For low values of $U\apprle1$, all
the excitations are above the gap. However, for $U\apprle1$, there
are sub-gap states, corresponding to collective modes. The structure
of the eigenvectors gives insight about the real-space structure of
the excitation. In Fig.~\ref{fig:ExSp_V2=00003D0.5}b) we show the
two-particle wave-function for a excitation with energy above the
gap. In this case, the wavefunction is well approximated by the product
between the two single-particle wavefunctions,
\[
\Psi_{X}^{eo}\left(r_{1},r_{2}\right)=\psi_{e}\left(r_{1}\right)\psi_{o}\left(r_{2}\right),
\]
with $e$ the index of an empty state, $o$ of an occupied state,
and $r_{1/2}$ the position of each particle. In Fig.~\ref{fig:ExSp_V2=00003D0.5}c)
we show a collective mode with energy below the gap, where  the real-space
representation corresponds to a bound state.

\begin{figure}
\begin{centering}
\includegraphics[width=0.98\columnwidth]{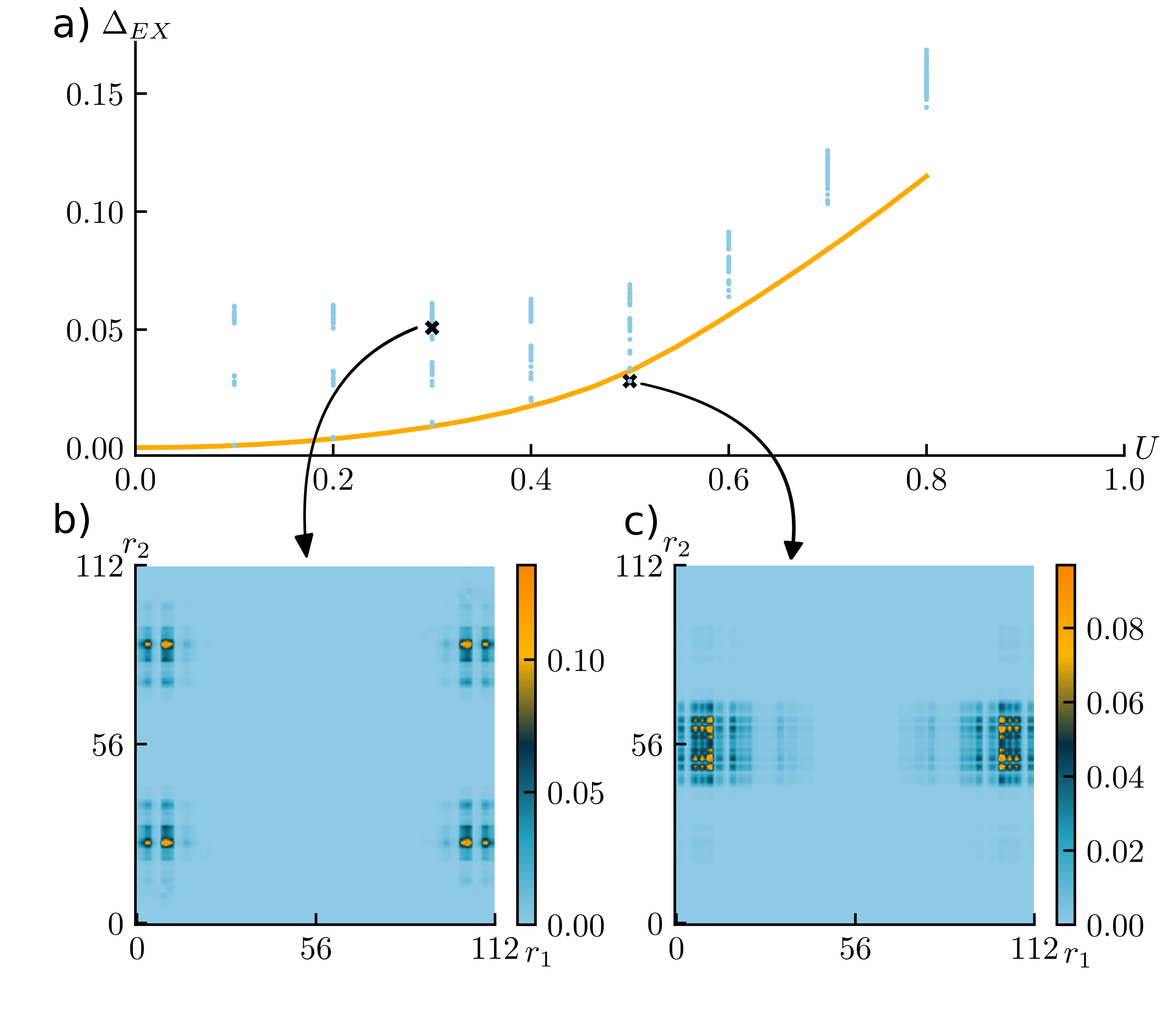}
\par\end{centering}
\caption{\label{fig:ExS_V2=00003D2.0}Excitation spectrum for the critical
regime, $V_{2}=2.0$ and system size $N=112$. a) Blue dots mark the
low-energy excitation energies and the orange line corresponds to
the mean-field gap. Real-space representation of an eigenvector for
an excitation with energy above the gap in b) and below the gap in
c).}
\end{figure}

Starting now in the critical regime, $V_{2}>1$, we may wander if
the quasifractal CDW has a characteristic excitation spectrum. If
the mean field ground state exhibits an incommensurate CDW (coming
from an incommensurate filling, far from half-filling), there is a
collective mode with zero excitation energy, described by the relative
shift of the wave in the underlying lattice, called the phason mode
. In the Appendix~\ref{sec:ICDW}, we show the excitation spectrum
for $V_{2}=0$, away from half-filling, where a zero-energy mode arises.
One would expect that these type of modes should appear in the quasifractal
regime of the CDW, but our results dismiss this possibility since
no zero energy mode (not even a sub-gap one) appear in the excitation
spectrum as can be seen in Fig.~\ref{fig:ExS_V2=00003D2.0}a). The
structure of the eigenvectors, shown in~\ref{fig:ExS_V2=00003D2.0}b)
and~\ref{fig:ExS_V2=00003D2.0}c) is simply the product of the two
single-particle wavefunctions, and the method does not give any new
information.

\subsection{Response functions}

\begin{figure}
\begin{centering}
\includegraphics[width=0.99\columnwidth]{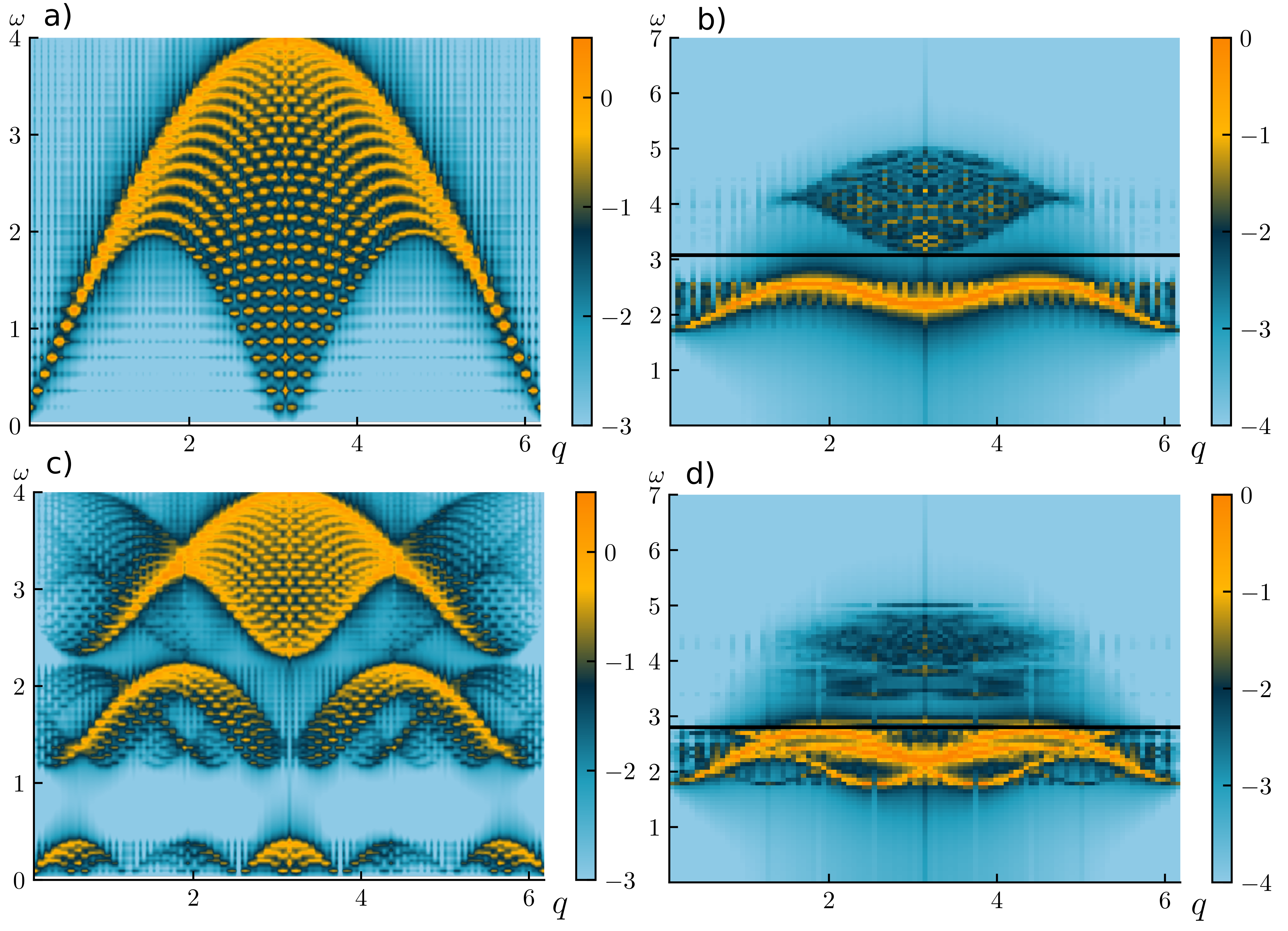}
\par\end{centering}
\caption{\label{fig:IChiU=00003D0V2=00003D0}Imaginary part of the charge response
function in the frequency-momentum domain in the non-interacting limit,
a) $V_{2}=0$ and b) $V_{2}=0.5$, and with interactions for $U=2$,
with $V_{2}=0$ in a) and $V_{2}=0.5$ in b).}
\end{figure}

The non-interacting charge susceptiblity, for $V_{2}=0$, is the analogue
of the Lindhard function to a lattice model, which we present in Fig.~\ref{fig:IChiU=00003D0V2=00003D0}a).
The zero-energy excitation at $\omega=0$ and $q=\pi$ shows the Fermi
surface nesting for that momentum and explains the instability of
the clean system to the $\pi$-CDW, when adding interactions. 

When introducing the quasiperiodic modulation, a narrow-band appears
in the single-particle spectrum, around the Fermi level, and two remote
bands are formed below and above. In Fig.~\ref{fig:IChiU=00003D0V2=00003D0}c),
three different energy regions can be distinguished, that are related
to excitations between those three bands. Focusing on the low-energy
excitations, the excitation spectrum exhibits many replicas of the
original response function, but centered at different momenta. The
values of the momenta for which an excitation is possible for $\omega=0,$
are precisely the ones that contribute to the CDW, $K_{n}=\left(\pi+2\pi\tau n\right)$,
explaining the mean field instability to the the quasiperiodic moir\'e
CDW. 

When  interactions are introduced, the collective modes dominate the
response. In Fig.~\ref{fig:IChiU=00003D0V2=00003D0}b), we show the
response at $U=2$ for $V_{2}=0$, where we notice that an excitation
band appear below the mean field gap (black line). With respect to
the modes in the particle-hole continuum, we notice that the largest
part correspond to $q=\pi$. In Fig.~\ref{fig:IChiU=00003D0V2=00003D0}b),
even though we have a finite $V_{2}=0.5$, and the sub-gap modes exhibits
a fine structure, the overall behavior remains the same.

\section{Conclusions \label{sec:Conclusions}}

We have determined the mean-field phase diagram of a 1D interacting
narrow-band moir\'e system, which exhibits extended and multifractal
critical single-particle states. We compared our results with the
exact phase diagram, obtained within the DMRG framework in Ref.~\citep{goncalves_incommensurability_2023},
to estabilish the effectiveness of the Hartree-Fock approximation
for this system. In the extended regime, we have found a Charge Density
Wave, with exponentially suppressed amplitude for any interaction
strength, a typical mean field result. The charge modulation exhibits
a finite number of wavevectors which is in agreement with the exact
result, although failing to capture the phase transition. In the critical
regime, the system is unstable to a charge density wave with a different
structure, where an extremely large number of wavevectors are present
when computing the Fourier transform of the density fluctuations,
diverging in the limit of vanishing interaction. This is in total
agreement with the exact calculation, which is a remarkable result
for a mean field approach, especially in 1D systems. These findings
support that the Hartree-Fock approximation can accurately describe
correlated phases of matter when the parent single-particle states
are multifractal and critical exhibiting delocalization in reciprocal
and real space. Furthermore, we used a real space implementation of
the time-dependent Hartree-Fock to study the excitation spectrum of
the system. We computed the spectrum for the extended and critical
regimes, where we found that for sufficiently large interaction strength,
sub-gap collective modes appear. With the excitation spectrum, we
were able to compute response functions, such as the charge susceptibility
to study the instabilities of the system. Studying the non-interacting
limit, we were able to observe the instabilities, at the mean field
level, to the different charge density waves that arise in the static
mean field calculations.

Our results are a significant step in the study of the interplay between
quasiperiodic systems and correlated phases, further demonstrating
that quasiperiodicity may stabilize novel interacting phases. Our
description establishes that a mean field approach is accurate for
systems exhibiting multifractal critical states. The recently studied
critical phases on the twisted bilayer graphene are a particular example
of a system where quasiperiodicity may induce novel correlated phases,
however its interplay is yet to be studied. We are confident that
a mean field approximation in such system should yield accurate results,
in particular due to the increased dimensionality, where the mean
field method, by itself, is expected to be more reliable.
\begin{acknowledgments}
The authors NS and EVC acknowledge FCT-Portugal through Grant No.
UID/04650 - Centro de Física das Universidades do Minho e do Porto.
 MG acknowledges partial support from Fundação para a Ciência e Tecnologia
(FCT-Portugal) through Grant No. UID/CTM/04540/2019. MG acknowledges
further support from FCT-Portugal through the Grant SFRH/BD/145152/2019.
\end{acknowledgments}

\appendix

\section{Time-dependent Hartree-Fock as a linear response theory\label{sec:Time-dependent-Hartree-Fock}}

Within the linear response theory, it is possible to obtain the response
of a system (expectation value of a certain operator, $\left\langle \hat{B}\right\rangle $,
subject to an external perturbation, $F(t)$, that couples to the
system's Hamiltonian through some operator $\hat{A}$. In this limit,
the response is completely governed by a generalized susceptibility,
\[
\chi_{BA}\left(t,t^{\prime},\boldsymbol{r},\boldsymbol{r}^{\prime}\right)=-\frac{i}{\hbar}\left\langle \left[B\left(\boldsymbol{r},t\right),A\left(\boldsymbol{r}^{\prime},t^{\prime}\right)\right]\right\rangle ,
\]
 as
\[
\left\langle B\right\rangle \left(\boldsymbol{r},t\right)=\left\langle B\right\rangle _{0}\left(\boldsymbol{r},t\right)+\int d\boldsymbol{r}^{\prime}\int_{-\infty}^{\infty}\chi_{BA}\left(\boldsymbol{r},\boldsymbol{r}^{\prime};t-t^{\prime}\right)F\left(t^{\prime}\right),
\]
where $\left\langle B\right\rangle _{0}\left(\boldsymbol{r},t\right)$
is the equilibrium expectation value. However, the exact many-body
ground state is not known, so an approximation must be employed. The
time-dependent Hartree-Fock, seen as a linear response theory is a
way to obtain the generalized susceptibility as a many-body perturbation
theory.

Since the response of any one-particle observable may be written as
\[
\left\langle B\right\rangle =\sum_{ab}B_{ab}\left\langle c_{a}^{\dagger}\left(t\right)c_{b}\left(t\right)\right\rangle ,
\]
where $a$ and $b$ are generic indices and $\left\langle c_{a}^{\dagger}\left(t\right)c_{b}\left(t\right)\right\rangle \equiv\rho_{ba}\left(t\right)$,
is the sistem's time-dependent reduced density matrix, whose time
evolution is calculated using Heisenberg's equation of motion. Working
on an arbitrary electronic basis $\left\{ \psi_{i}\right\} $, defined
by its creation operators, $\ket{\psi_{i}}=d_{i}^{\dagger}\ket{0}$,
the many-body Hamiltonian reads
\[
H=H_{0}+H_{\text{int}}+\hat{A}F(t),
\]
with 
\begin{align*}
H_{0} & =\sum_{ij}t_{ij}d_{i}^{\dagger}d_{j}\\
H_{\text{int}} & =\sum_{ijkl}U_{kl}^{ij}d_{i}^{\dagger}d_{j}^{\dagger}d_{k}d_{l}\\
\hat{A} & =\sum_{ij}A_{ij}d_{i}^{\dagger}d_{j},
\end{align*}
where $A_{ij}$ are the matrix elements of the time-dependent perturbation
that drives the system out of equilibrium inducing transition from
a given state $\psi_{j}$ to another state $\psi_{i}$. The time evolution
of the reduced density matrix is given by 
\[
\frac{d}{dt}\rho_{ba}=\left\langle d_{a}^{\dagger}\frac{d}{dt}d_{b}\right\rangle +\left\langle \frac{d}{dt}d_{a}^{\dagger}d_{b}\right\rangle ,
\]
where the time evolution is given by Heisenberg's equation of motion
\[
\frac{d}{dt}\mathcal{O}(t)=\frac{i}{\hbar}\left[H,\mathcal{O}(t)\right].
\]
Then, the time evolution of the reduced density matrix is simply
\[
\frac{d}{dt}\rho_{ba}=\frac{i}{\hbar}\left\langle d_{a}^{\dagger}\left[H,d_{b}\right]\right\rangle +\frac{i}{\hbar}\left\langle \left[H,d_{a}^{\dagger}\right]d_{b}\right\rangle .
\]
The first commutator is given by 
\begin{align*}
\left[H,d_{b}\right] & =-\sum_{j}\left(t_{bj}+A_{bj}F(t)\right)d_{j}\\
 & +\sum_{jkl}\left(U_{kl}^{jb}-U_{kl}^{bj}\right)d_{j}^{\dagger}d_{k}d_{l},
\end{align*}
and the second one by
\begin{align*}
\left[H,d_{a}^{\dagger}\right] & =\sum_{i}\left(t_{ia}+A_{ia}F(t)\right)d_{i}^{\dagger}\\
 & +\sum_{ijk}\left(U_{ka}^{ij}-U_{ak}^{ij}\right)d_{i}^{\dagger}d_{j}^{\dagger}d_{k}.
\end{align*}
Therefore, the reduced density matrix evolves in time as
\begin{align*}
-i\hbar\frac{d}{dt}\rho_{ba}= & -\sum_{j}\left(t_{bj}+A_{bj}F(t)\right)\rho_{ja}\\
 & +\sum_{i}\left(t_{ia}+A_{ia}F(t)\right)\rho_{bi}\\
 & +\sum_{jkl}\left(U_{kl}^{jb}-U_{kl}^{bj}\right)\left\langle d_{a}^{\dagger}d_{j}^{\dagger}d_{k}d_{l}\right\rangle \\
 & +\sum_{ijk}\left(U_{ka}^{ij}-U_{ak}^{ij}\right)\left\langle d_{i}^{\dagger}d_{j}^{\dagger}d_{k}d_{b}\right\rangle .
\end{align*}
The tdHF approximation assumes that, at all times, the ground state
behaves as a Slater determinant, so we employ a Wick's decoupling
in the four-fermion average value as
\[
\left\langle d_{a}^{\dagger}d_{j}^{\dagger}d_{k}d_{l}\right\rangle =\left\langle d_{a}^{\dagger}d_{l}\right\rangle \left\langle d_{j}^{\dagger}d_{k}\right\rangle -\left\langle d_{a}^{\dagger}d_{k}\right\rangle \left\langle d_{j}^{\dagger}d_{l}\right\rangle ,
\]
where we discarded anomalous terms. Applying this approximation, we
arrive at 
\begin{align*}
-i\hbar\frac{d}{dt}\rho_{ba}\left(t\right) & =\sum_{i}\left(t_{ia}+A_{ia}F(t)\right)\rho_{bi}\\
 & -\sum_{j}\left(t_{bj}+A_{bj}F(t)\right)\rho_{ja}\\
 & +\sum_{l}\Sigma_{bl}\left[\boldsymbol{\rho}\right]\rho_{la}+\rho_{bi}\Sigma_{ia}\left[\boldsymbol{\rho}\right],
\end{align*}
where 
\begin{align*}
\Sigma_{bl}\left[\boldsymbol{\rho}\right] & =\sum_{kl}\left(U_{kl}^{jb}+U_{lk}^{bj}-U_{kl}^{bj}-U_{lk}^{jb}\right)\rho_{kj},
\end{align*}
is defined as a Hartree-Fock self-energy. Since we are interested
only in the linear response regime, let us introduce a formal expansion
of the reduced density matrix,
\[
\boldsymbol{\rho}\left(t\right)=\sum_{n}\boldsymbol{\rho}^{(n)}(t),
\]
where $n$ indicates the order of the perturbation. Since the single-particle
elements, $t_{ia}/t_{bj}$, has its origin in a static mean-field
approach, we have to subtract the equilibrium density matrix in the
self-energy term to avoid double counting as 
\[
\Sigma\left[\boldsymbol{\rho}\right]\to\Sigma\left[\boldsymbol{\rho}-\boldsymbol{\rho}^{(0)}\right].
\]
Collecting terms that are linear in the perturbation, we obtain
\begin{align*}
i\hbar\frac{d}{dt}\rho_{ba}^{(1)}\left(t\right)= & \sum_{i}\left(t_{bi}\rho_{ia}^{(1)}-\rho_{bi}^{(1)}t_{ia}\right)\\
+ & \sum_{i}\left(A_{bi}\rho_{ia}^{(0)}-\rho_{bi}^{(0)}A_{ia}\right)F(t)\\
+ & \sum_{ijk}\left(U_{ki}^{bj}+U_{ik}^{jb}-U_{ki}^{jb}-U_{ik}^{bj}\right)\rho_{kj}^{(1)}\rho_{ia}^{(0)}\\
- & \sum_{ijk}\rho_{bi}^{(0)}\left(U_{ka}^{ij}+U_{ak}^{ji}-U_{ak}^{ij}-U_{ka}^{ji}\right)\rho_{kj}^{(1)}.
\end{align*}
Performing a Fourier transform in tim we may write the equation as
\begin{equation}
\sum_{cd}\left(\hbar\omega\delta_{c\alpha}\delta_{\beta d}-H_{\alpha\beta}^{cd}\right)\rho_{cd}^{(1)}\left(\omega\right)=J_{\alpha\beta}\left(\omega\right),\label{eq:LRs}
\end{equation}
where 
\begin{align*}
H_{\alpha\beta}^{cd} & =t_{\alpha c}\delta_{d\beta}-t_{d\beta}\delta_{\alpha c}\\
 & +\sum_{e}\left(U_{ce}^{\alpha d}+U_{ec}^{d\alpha}-U_{ce}^{d\alpha}-U_{ec}^{\alpha d}\right)\rho_{e\beta}^{(0)}\\
 & -\sum_{e}\rho_{\alpha e}^{(0)}\left(U_{c\beta}^{ed}+U_{\beta c}^{de}-U_{\beta c}^{ed}-U_{c\beta}^{de}\right),
\end{align*}
and 
\[
J_{\alpha\beta}\left(\omega\right)=\sum_{e}\left(A_{\alpha e}\rho_{e\beta}^{(0)}-\rho_{\alpha e}^{(0)}A_{e\beta}\right)F(\omega).
\]
Choosing the basis that diagonalizes the single particle Hamiltonian,
in which $\rho_{\alpha\beta}^{(0)}=f_{\alpha}\delta_{\alpha\beta}$,
with $f_{\alpha}$ the Fermi occupation factors and $t_{\alpha\beta}=E_{\alpha}^{(MF)}\delta_{\alpha\beta},$with
$E_{\alpha}^{(MF)}$ the $\alpha-$th eigenenergy of the mean field
Hamiltonian, we may write
\begin{align}
H_{\alpha\beta}^{cd} & =\left(E_{\alpha}^{(\text{MF})}-E_{\beta}^{(\text{MF})}\right)\delta_{\alpha c}\delta_{\beta d}\nonumber \\
 & +\left(U_{c\beta}^{\alpha d}+U_{\beta c}^{d\alpha}-U_{c\beta}^{d\alpha}-U_{\beta c}^{\alpha d}\right)\left(f_{\beta}-f_{\alpha}\right),\label{eq:H_abcd}
\end{align}
and
\[
J_{\alpha\beta}\left(\omega\right)=\left(f_{\beta}-f_{\alpha}\right)A_{\alpha\beta}F(\omega).
\]
Now, considering the limit of zero temperature, the states are occupied
or empty, so it is possible to rewrite Eq. \eqref{eq:LRs} by splitting
each index in those two set as
\[
\left(\begin{array}{cccc}
H_{oo}^{o^{\prime}o^{\prime}} & H_{oo}^{o^{\prime}e^{\prime}} & H_{oo}^{e^{\prime}o^{\prime}} & H_{oo}^{e^{\prime}e^{\prime}}\\
H_{oe}^{o^{\prime}o^{\prime}} & H_{oe}^{o^{\prime}e^{\prime}} & H_{oe}^{e^{\prime}o^{\prime}} & H_{oe}^{e^{\prime}e^{\prime}}\\
H_{eo}^{o^{\prime}o^{\prime}} & H_{eo}^{o^{\prime}e^{\prime}} & H_{eo}^{e^{\prime}o^{\prime}} & H_{eo}^{e^{\prime}e^{\prime}}\\
H_{ee}^{o^{\prime}o^{\prime}} & H_{ee}^{o^{\prime}e^{\prime}} & H_{ee}^{e^{\prime}o^{\prime}} & H_{ee}^{e^{\prime}e^{\prime}}
\end{array}\right)\left(\begin{array}{c}
\rho_{o^{\prime}o^{\prime}}^{(1)}\\
\rho_{o^{\prime}e^{\prime}}^{(1)}\\
\rho_{e^{\prime}o^{\prime}}^{(1)}\\
\rho_{e^{\prime}e^{\prime}}^{(1)}
\end{array}\right)=\left(\begin{array}{c}
J_{oo}\left(\omega\right)\\
J_{oe}\left(\omega\right)\\
J_{eo}\left(\omega\right)\\
J_{ee}\left(\omega\right)
\end{array}\right).
\]
Due to zero temperature limit and the Hermitian property of the interaction
matrix elements, only the following matrix elements survive and the
equation reduces only to electron-hole excitations.
\[
\left(\begin{array}{cc}
R_{oe}^{o^{\prime}e^{\prime}} & C_{oe}^{e^{\prime}o^{\prime}}\\
C_{eo}^{o^{\prime}e^{\prime}} & R_{eo}^{e^{\prime}o^{\prime}}
\end{array}\right)\left(\begin{array}{c}
\rho_{o^{\prime}e^{\prime}}^{(1)}\\
\rho_{e^{\prime}o^{\prime}}^{(1)}
\end{array}\right)=\left(\begin{array}{c}
J_{oe}\left(\omega\right)\\
J_{eo}\left(\omega\right)
\end{array}\right),
\]
where we have defined 
\begin{align*}
R_{oe}^{o^{\prime}e^{\prime}} & =\left(E_{o}-E_{e}\right)\delta_{ee^{\prime}}\delta_{oo^{\prime}}+\left(U_{o^{\prime}e}^{oe^{\prime}}+U_{eo^{\prime}}^{e^{\prime}o}-U_{o^{\prime}e}^{e^{\prime}o}-U_{eo^{\prime}}^{oe^{\prime}}\right)\\
C_{oe}^{e^{\prime}o^{\prime}} & =U_{e^{\prime}e}^{oo^{\prime}}+U_{ee^{\prime}}^{o^{\prime}o}-U_{e^{\prime}e}^{o^{\prime}o}-U_{ee^{\prime}}^{oo^{\prime}}.
\end{align*}
The correction to the observable, $\delta\left\langle B\right\rangle \left(\omega\right)$
is thus given by 
\begin{align*}
\delta\left\langle B\right\rangle \left(\omega\right) & =\text{Tr}\left[\rho^{(1)}\left(\omega\right)B\right]\\
 & =\left[\boldsymbol{B}_{eo}\,\,\,\boldsymbol{B}_{oe}\right]\left[\begin{array}{cc}
\hbar\omega\boldsymbol{I}-\boldsymbol{R} & -\boldsymbol{C}\\
\boldsymbol{C}^{\dagger} & \hbar\omega\boldsymbol{I}+\boldsymbol{R}^{*}
\end{array}\right]^{-1}\left[\begin{array}{c}
\boldsymbol{A}_{eo}\\
-\boldsymbol{A}_{oe}
\end{array}\right]
\end{align*}

\section{Collective Modes of the 1D Hubbard Model\label{sec:Hubbardf}}

\begin{figure}[t]
\centering{}\includegraphics[width=0.99\columnwidth]{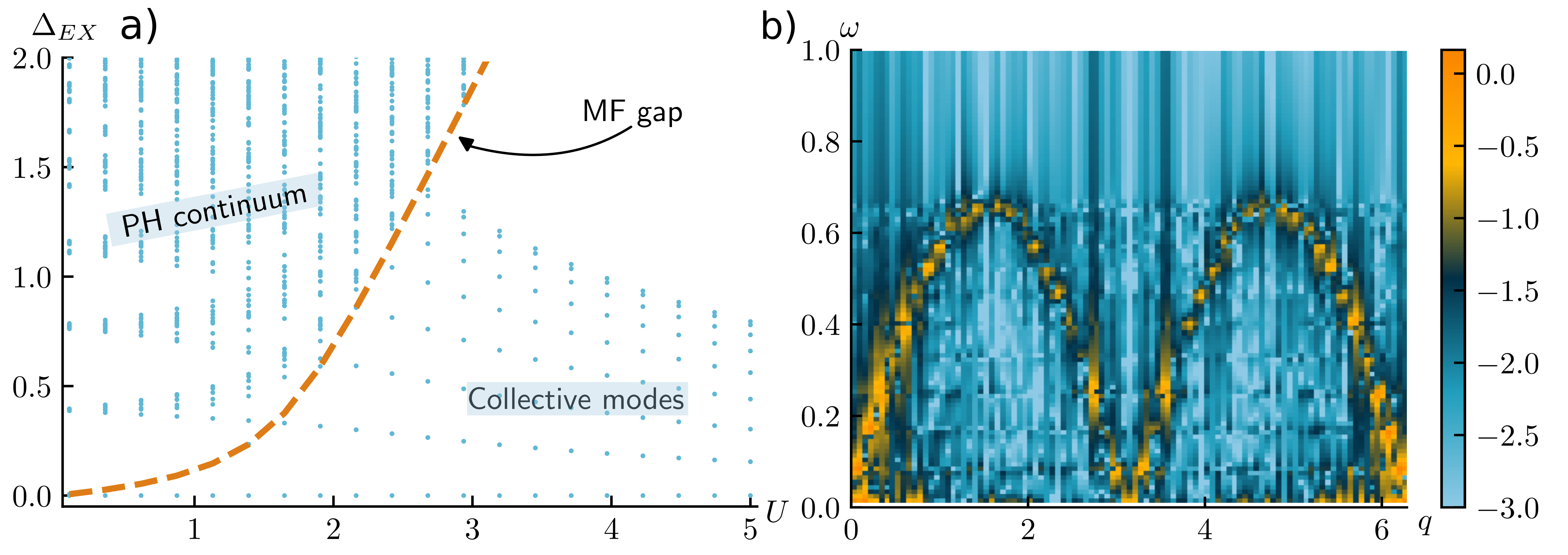}\caption{\label{fig:Hubbard}a) Excitation spectrum for the 1D Hubbard model
for a finite system with size $N=56$. The orange line marks the mean
field gap, the points above the gap correspond to the particle-hole
(PH) continuum and the ones below to the collective modes. Note that
a zero energy mode appears for every interaction strength. b) Imaginary
part (logarithmic scale) of the transverse spin susceptibility, for
$U=5$, corresponding to the momentum-resolved excitation spectrum.
The orange lines define the spin-wave spectrum, $\omega(q)=\frac{4}{U}\left|\sin(q)\right|$,
with the Goldstone mode at $q\to0$.}
\end{figure}

The Hamiltonian for the 1D Hubbard model reads
\begin{equation}
H_{\text{Hub}}=-t\sum_{i\sigma}c_{i\sigma}^{\dagger}c_{i+1\sigma}+\text{h.c.}+U\sum_{n}n_{i\uparrow}n_{i\downarrow},\label{eq:H_hub}
\end{equation}
where $c_{i\sigma}^{\dagger}$ creates an electron of spin $\sigma$
at site $i$, $n_{i\sigma}=c_{i\sigma}^{\dagger}c_{i\sigma}$ is the
number operator and $U$ is the Hubbard repulsive term. We perform
a mean field decoupling with a mean field Hamiltonian
\begin{align*}
H_{\text{MF}} & =-t\sum_{i\sigma}c_{i\sigma}^{\dagger}c_{i+1\sigma}+\text{h.c.}+U\sum_{i\sigma}\left\langle n_{i-\sigma}\right\rangle n_{i\sigma}.\\
 & -U\sum_{i\sigma}\left\langle c_{i\uparrow}^{\dagger}c_{i\downarrow}\right\rangle c_{i\downarrow}^{\dagger}c_{i\downarrow}+\text{h.c.}.
\end{align*}
We did not consider the Fock terms as they correspond only to a rotation
in the magnetic polarization. Solving the self-consistent mean field
equations we obtain an anti-ferromagnetic ground state with exponentially
suppressend magnetization. However, this polarization points in the
$z-$direction, since it is the direction of the quantization, but
a ground state with the same internal energy. Furthermore, the Hamiltonian
in Eq. \eqref{eq:H_hub} exhibits a rotation symmetry (SO(3)). A Goldstone
mode is expected, since the mean field ground state has a lower degree
of symmetry compared to the Hamiltonian. Using the tdHF, we were able
to obtain the excitation spectrum of the Hubbard model. In particular,
a zero-energy mode is always found. In Fig.~\ref{fig:Hubbard}a),
we present the low-energy sector of the excitations. To obtain a momentum-resolved
description of the excitations we focus our attention in the (transverse)
spin--spin response function. In Fig.~\ref{fig:Hubbard}b) we show
the imaginary part of the response function for $U=5$. We note that
the excitation spectrum correspond to 
\[
\omega(q)=\frac{4}{U}\left|\sin(q)\right|,
\]
which is precisely the spin-wave spectrum for the antiferromagnetic
case. In particular, we note that, when $q\to0$, the excitation energy
$\omega(q)\to0$, corresponding to the Goldstone mode. With this example,
we show, that our method is well-capable for the description of collective
modes in fermionic systems, with a real-space description.

\section{Incommensurate Charge Density Wave\label{sec:ICDW}}

\begin{figure}
\begin{centering}
\includegraphics[width=0.98\columnwidth]{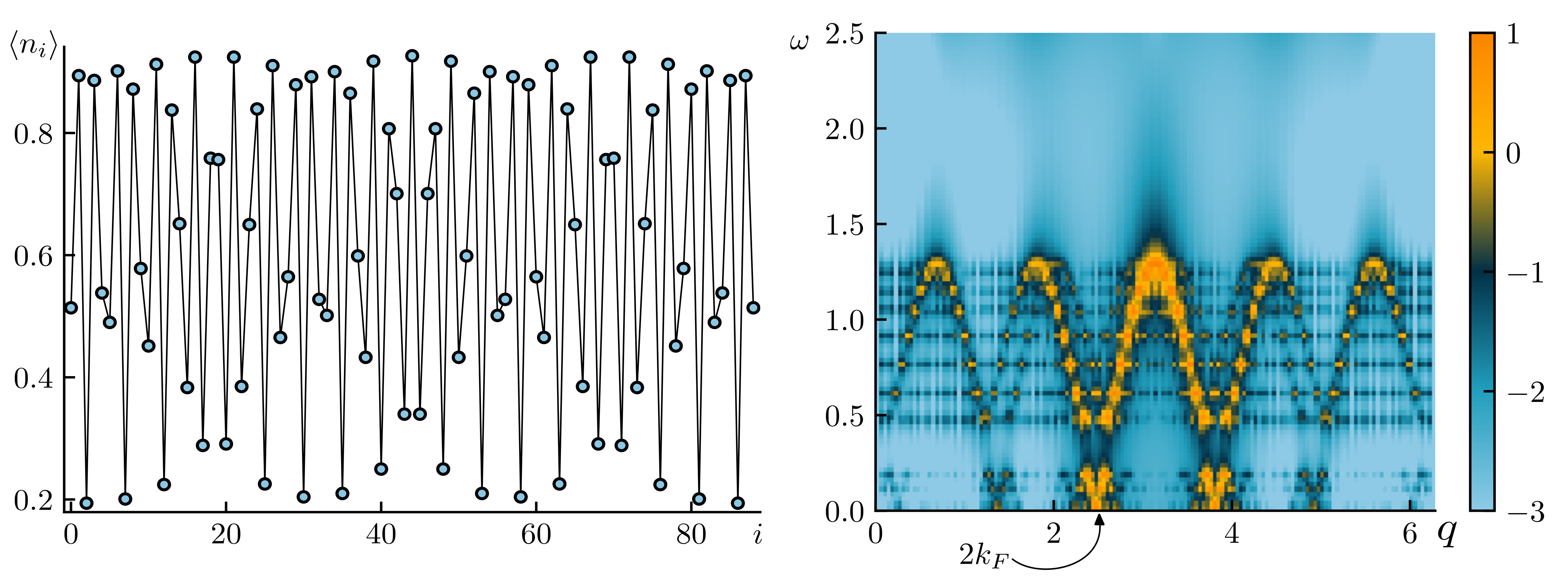}
\par\end{centering}
\caption{\label{fig:ICDW}a) Charge density modulation for $U=5$ for a system
size $N=89$ and filling factor, $\nu=\frac{55}{89}$. b) Imaginary
part of the charge response function for the same system parameters.}
\end{figure}

We now consider the Hamiltonian of Eq.~\ref{eq:H_int} for $V_{2}=0$
but away from half-filling. On this case, we obtain the so-called
Incommensurate Charge Density Wave (I-CDW), shown in Fig.~\ref{fig:ICDW}a)
with a periodicity equal to the system size. There are two possible
collective modes of this charge structure, the phason and the amplitudon.
The phason one, a gapless mode, corresponds to a slide of the density
wave with respect to the underlying lattice. The amplitudon, with
a finite energy gap, is a oscillation of the amplitude of the density.
We compute the charge response function for this system, with $U=5$
as a function of the frequency and domain, as shown in Fig.~\ref{fig:ICDW}b)
where we have the low-energy excitations. In particular, we obtain
the zero energy modes, at $q=2k_{F}$, corresponding to the modulation
of the phason mode.

\bibliographystyle{apsrev}
\bibliography{refs}

\begin{thebibliography}{52}
\expandafter\ifx\csname natexlab\endcsname\relax\def\natexlab#1{#1}\fi
\expandafter\ifx\csname bibnamefont\endcsname\relax
  \def\bibnamefont#1{#1}\fi
\expandafter\ifx\csname bibfnamefont\endcsname\relax
  \def\bibfnamefont#1{#1}\fi
\expandafter\ifx\csname citenamefont\endcsname\relax
  \def\citenamefont#1{#1}\fi
\expandafter\ifx\csname url\endcsname\relax
  \def\url#1{\texttt{#1}}\fi
\expandafter\ifx\csname urlprefix\endcsname\relax\def\urlprefix{URL }\fi
\providecommand{\bibinfo}[2]{#2}
\providecommand{\eprint}[2][]{\url{#2}}

\bibitem[{\citenamefont{Wang et~al.}(2020{\natexlab{a}})\citenamefont{Wang,
  Shih, Ghiotto, Xian, Rhodes, Tan, Claassen, Kennes, Bai, Kim
  et~al.}}]{wang_correlated_2020}
\bibinfo{author}{\bibfnamefont{L.}~\bibnamefont{Wang}},
  \bibinfo{author}{\bibfnamefont{E.-M.} \bibnamefont{Shih}},
  \bibinfo{author}{\bibfnamefont{A.}~\bibnamefont{Ghiotto}},
  \bibinfo{author}{\bibfnamefont{L.}~\bibnamefont{Xian}},
  \bibinfo{author}{\bibfnamefont{D.~A.} \bibnamefont{Rhodes}},
  \bibinfo{author}{\bibfnamefont{C.}~\bibnamefont{Tan}},
  \bibinfo{author}{\bibfnamefont{M.}~\bibnamefont{Claassen}},
  \bibinfo{author}{\bibfnamefont{D.~M.} \bibnamefont{Kennes}},
  \bibinfo{author}{\bibfnamefont{Y.}~\bibnamefont{Bai}},
  \bibinfo{author}{\bibfnamefont{B.}~\bibnamefont{Kim}}, \bibnamefont{et~al.},
  \bibinfo{journal}{Nature Materials} \textbf{\bibinfo{volume}{19}},
  \bibinfo{pages}{861} (\bibinfo{year}{2020}{\natexlab{a}}), ISSN
  \bibinfo{issn}{1476-4660}.

\bibitem[{\citenamefont{Ghiotto et~al.}(2021)\citenamefont{Ghiotto, Shih,
  Pereira, Rhodes, Kim, Zang, Millis, Watanabe, Taniguchi, Hone
  et~al.}}]{ghiotto_quantum_2021}
\bibinfo{author}{\bibfnamefont{A.}~\bibnamefont{Ghiotto}},
  \bibinfo{author}{\bibfnamefont{E.-M.} \bibnamefont{Shih}},
  \bibinfo{author}{\bibfnamefont{G.~S. S.~G.} \bibnamefont{Pereira}},
  \bibinfo{author}{\bibfnamefont{D.~A.} \bibnamefont{Rhodes}},
  \bibinfo{author}{\bibfnamefont{B.}~\bibnamefont{Kim}},
  \bibinfo{author}{\bibfnamefont{J.}~\bibnamefont{Zang}},
  \bibinfo{author}{\bibfnamefont{A.~J.} \bibnamefont{Millis}},
  \bibinfo{author}{\bibfnamefont{K.}~\bibnamefont{Watanabe}},
  \bibinfo{author}{\bibfnamefont{T.}~\bibnamefont{Taniguchi}},
  \bibinfo{author}{\bibfnamefont{J.~C.} \bibnamefont{Hone}},
  \bibnamefont{et~al.}, \bibinfo{journal}{Nature}
  \textbf{\bibinfo{volume}{597}}, \bibinfo{pages}{345} (\bibinfo{year}{2021}),
  ISSN \bibinfo{issn}{1476-4687}.

\bibitem[{\citenamefont{Xu et~al.}(2022)\citenamefont{Xu, Kang, Watanabe,
  Taniguchi, Mak, and Shan}}]{xu_tunable_2022}
\bibinfo{author}{\bibfnamefont{Y.}~\bibnamefont{Xu}},
  \bibinfo{author}{\bibfnamefont{K.}~\bibnamefont{Kang}},
  \bibinfo{author}{\bibfnamefont{K.}~\bibnamefont{Watanabe}},
  \bibinfo{author}{\bibfnamefont{T.}~\bibnamefont{Taniguchi}},
  \bibinfo{author}{\bibfnamefont{K.~F.} \bibnamefont{Mak}}, \bibnamefont{and}
  \bibinfo{author}{\bibfnamefont{J.}~\bibnamefont{Shan}},
  \bibinfo{journal}{Nature Nanotechnology} \textbf{\bibinfo{volume}{17}},
  \bibinfo{pages}{934} (\bibinfo{year}{2022}), ISSN \bibinfo{issn}{1748-3395}.

\bibitem[{\citenamefont{Li et~al.}(2024)\citenamefont{Li, Xiang, Naik, Kim, Li,
  Sailus, Banerjee, Taniguchi, Watanabe, Tongay et~al.}}]{li_imaging_2024}
\bibinfo{author}{\bibfnamefont{H.}~\bibnamefont{Li}},
  \bibinfo{author}{\bibfnamefont{Z.}~\bibnamefont{Xiang}},
  \bibinfo{author}{\bibfnamefont{M.~H.} \bibnamefont{Naik}},
  \bibinfo{author}{\bibfnamefont{W.}~\bibnamefont{Kim}},
  \bibinfo{author}{\bibfnamefont{Z.}~\bibnamefont{Li}},
  \bibinfo{author}{\bibfnamefont{R.}~\bibnamefont{Sailus}},
  \bibinfo{author}{\bibfnamefont{R.}~\bibnamefont{Banerjee}},
  \bibinfo{author}{\bibfnamefont{T.}~\bibnamefont{Taniguchi}},
  \bibinfo{author}{\bibfnamefont{K.}~\bibnamefont{Watanabe}},
  \bibinfo{author}{\bibfnamefont{S.}~\bibnamefont{Tongay}},
  \bibnamefont{et~al.}, \bibinfo{journal}{Nature materials}
  \textbf{\bibinfo{volume}{23}}, \bibinfo{pages}{633} (\bibinfo{year}{2024}),
  ISSN \bibinfo{issn}{1476-1122},
  \urlprefix\url{http://www.scopus.com/inward/record.url?scp=85181250120&partnerID=8YFLogxK}.

\bibitem[{\citenamefont{Kim et~al.}(2016)\citenamefont{Kim, Yankowitz,
  Fallahazad, Kang, Movva, Huang, Larentis, Corbet, Taniguchi, Watanabe
  et~al.}}]{kim_van_2016}
\bibinfo{author}{\bibfnamefont{K.}~\bibnamefont{Kim}},
  \bibinfo{author}{\bibfnamefont{M.}~\bibnamefont{Yankowitz}},
  \bibinfo{author}{\bibfnamefont{B.}~\bibnamefont{Fallahazad}},
  \bibinfo{author}{\bibfnamefont{S.}~\bibnamefont{Kang}},
  \bibinfo{author}{\bibfnamefont{H.~C.} \bibnamefont{Movva}},
  \bibinfo{author}{\bibfnamefont{S.}~\bibnamefont{Huang}},
  \bibinfo{author}{\bibfnamefont{S.}~\bibnamefont{Larentis}},
  \bibinfo{author}{\bibfnamefont{C.~M.} \bibnamefont{Corbet}},
  \bibinfo{author}{\bibfnamefont{T.}~\bibnamefont{Taniguchi}},
  \bibinfo{author}{\bibfnamefont{K.}~\bibnamefont{Watanabe}},
  \bibnamefont{et~al.}, \bibinfo{journal}{Nano letters}
  \textbf{\bibinfo{volume}{16}}, \bibinfo{pages}{1989} (\bibinfo{year}{2016}),
  ISSN \bibinfo{issn}{1530-6992}, \bibinfo{note}{publisher: Nano Lett},
  \urlprefix\url{https://pubmed.ncbi.nlm.nih.gov/26859527/}.

\bibitem[{\citenamefont{Cao et~al.}(2018{\natexlab{a}})\citenamefont{Cao,
  Fatemi, Fang, Watanabe, Taniguchi, Kaxiras, and
  Jarillo-Herrero}}]{cao_unconventional_2018}
\bibinfo{author}{\bibfnamefont{Y.}~\bibnamefont{Cao}},
  \bibinfo{author}{\bibfnamefont{V.}~\bibnamefont{Fatemi}},
  \bibinfo{author}{\bibfnamefont{S.}~\bibnamefont{Fang}},
  \bibinfo{author}{\bibfnamefont{K.}~\bibnamefont{Watanabe}},
  \bibinfo{author}{\bibfnamefont{T.}~\bibnamefont{Taniguchi}},
  \bibinfo{author}{\bibfnamefont{E.}~\bibnamefont{Kaxiras}}, \bibnamefont{and}
  \bibinfo{author}{\bibfnamefont{P.}~\bibnamefont{Jarillo-Herrero}},
  \bibinfo{journal}{Nature 2018 556:7699} \textbf{\bibinfo{volume}{556}},
  \bibinfo{pages}{43} (\bibinfo{year}{2018}{\natexlab{a}}), ISSN
  \bibinfo{issn}{1476-4687}, \bibinfo{note}{publisher: Nature Publishing
  Group}, \urlprefix\url{https://www.nature.com/articles/nature26160}.

\bibitem[{\citenamefont{Cao et~al.}(2018{\natexlab{b}})\citenamefont{Cao,
  Fatemi, Demir, Fang, Tomarken, Luo, Sanchez-Yamagishi, Watanabe, Taniguchi,
  Kaxiras et~al.}}]{cao_correlated_2018}
\bibinfo{author}{\bibfnamefont{Y.}~\bibnamefont{Cao}},
  \bibinfo{author}{\bibfnamefont{V.}~\bibnamefont{Fatemi}},
  \bibinfo{author}{\bibfnamefont{A.}~\bibnamefont{Demir}},
  \bibinfo{author}{\bibfnamefont{S.}~\bibnamefont{Fang}},
  \bibinfo{author}{\bibfnamefont{S.~L.} \bibnamefont{Tomarken}},
  \bibinfo{author}{\bibfnamefont{J.~Y.} \bibnamefont{Luo}},
  \bibinfo{author}{\bibfnamefont{J.~D.} \bibnamefont{Sanchez-Yamagishi}},
  \bibinfo{author}{\bibfnamefont{K.}~\bibnamefont{Watanabe}},
  \bibinfo{author}{\bibfnamefont{T.}~\bibnamefont{Taniguchi}},
  \bibinfo{author}{\bibfnamefont{E.}~\bibnamefont{Kaxiras}},
  \bibnamefont{et~al.}, \bibinfo{journal}{Nature 2018 556:7699}
  \textbf{\bibinfo{volume}{556}}, \bibinfo{pages}{80}
  (\bibinfo{year}{2018}{\natexlab{b}}), ISSN \bibinfo{issn}{1476-4687},
  \bibinfo{note}{publisher: Nature Publishing Group \_eprint: 1802.00553},
  \urlprefix\url{https://www.nature.com/articles/nature26154}.

\bibitem[{\citenamefont{Nuckolls et~al.}(2020)\citenamefont{Nuckolls, Oh, Wong,
  Lian, Watanabe, Taniguchi, Bernevig, and Yazdani}}]{nuckolls_strongly_2020}
\bibinfo{author}{\bibfnamefont{K.~P.} \bibnamefont{Nuckolls}},
  \bibinfo{author}{\bibfnamefont{M.}~\bibnamefont{Oh}},
  \bibinfo{author}{\bibfnamefont{D.}~\bibnamefont{Wong}},
  \bibinfo{author}{\bibfnamefont{B.}~\bibnamefont{Lian}},
  \bibinfo{author}{\bibfnamefont{K.}~\bibnamefont{Watanabe}},
  \bibinfo{author}{\bibfnamefont{T.}~\bibnamefont{Taniguchi}},
  \bibinfo{author}{\bibfnamefont{B.~A.} \bibnamefont{Bernevig}},
  \bibnamefont{and} \bibinfo{author}{\bibfnamefont{A.}~\bibnamefont{Yazdani}},
  \bibinfo{journal}{Nature} \textbf{\bibinfo{volume}{588}},
  \bibinfo{pages}{610} (\bibinfo{year}{2020}), ISSN \bibinfo{issn}{1476-4687}.

\bibitem[{\citenamefont{Lopes Dos~Santos et~al.}(2012)\citenamefont{Lopes
  Dos~Santos, Peres, and Castro~Neto}}]{lopes_dos_santos_continuum_2012}
\bibinfo{author}{\bibfnamefont{J.~M.} \bibnamefont{Lopes Dos~Santos}},
  \bibinfo{author}{\bibfnamefont{N.~M.} \bibnamefont{Peres}}, \bibnamefont{and}
  \bibinfo{author}{\bibfnamefont{A.~H.} \bibnamefont{Castro~Neto}},
  \bibinfo{journal}{Physical Review B - Condensed Matter and Materials Physics}
  \textbf{\bibinfo{volume}{86}}, \bibinfo{pages}{155449}
  (\bibinfo{year}{2012}), ISSN \bibinfo{issn}{10980121},
  \bibinfo{note}{publisher: American Physical Society},
  \urlprefix\url{https://journals.aps.org/prb/abstract/10.1103/PhysRevB.86.155449}.

\bibitem[{\citenamefont{Bistritzer and
  MacDonald}(2011)}]{bistritzer_moire_2011}
\bibinfo{author}{\bibfnamefont{R.}~\bibnamefont{Bistritzer}} \bibnamefont{and}
  \bibinfo{author}{\bibfnamefont{A.~H.} \bibnamefont{MacDonald}},
  \bibinfo{journal}{Proceedings of the National Academy of Sciences of the
  United States of America} \textbf{\bibinfo{volume}{108}},
  \bibinfo{pages}{12233} (\bibinfo{year}{2011}), ISSN \bibinfo{issn}{00278424},
  \bibinfo{note}{iSBN: 1108174108 \_eprint: 1009.4203}.

\bibitem[{\citenamefont{Gonçalves et~al.}(2021)\citenamefont{Gonçalves,
  Olyaei, Amorim, Mondaini, Ribeiro, and
  Castro}}]{goncalves_incommensurability-induced_2021}
\bibinfo{author}{\bibfnamefont{M.}~\bibnamefont{Gonçalves}},
  \bibinfo{author}{\bibfnamefont{H.~Z.} \bibnamefont{Olyaei}},
  \bibinfo{author}{\bibfnamefont{B.}~\bibnamefont{Amorim}},
  \bibinfo{author}{\bibfnamefont{R.}~\bibnamefont{Mondaini}},
  \bibinfo{author}{\bibfnamefont{P.}~\bibnamefont{Ribeiro}}, \bibnamefont{and}
  \bibinfo{author}{\bibfnamefont{E.~V.} \bibnamefont{Castro}},
  \bibinfo{journal}{2D Materials} \textbf{\bibinfo{volume}{9}},
  \bibinfo{pages}{011001} (\bibinfo{year}{2021}), \bibinfo{note}{publisher: IOP
  Publishing}, \urlprefix\url{https://dx.doi.org/10.1088/2053-1583/ac3259}.

\bibitem[{\citenamefont{Szab\'{o} and Schneider}(2020)}]{szabo_mixed_2020}
\bibinfo{author}{\bibfnamefont{A.}~\bibnamefont{Szab\'{o}}} \bibnamefont{and}
  \bibinfo{author}{\bibfnamefont{U.}~\bibnamefont{Schneider}},
  \bibinfo{journal}{Physical Review B} \textbf{\bibinfo{volume}{101}},
  \bibinfo{pages}{014205} (\bibinfo{year}{2020}), \bibinfo{note}{publisher:
  American Physical Society},
  \urlprefix\url{https://link.aps.org/doi/10.1103/PhysRevB.101.014205}.

\bibitem[{\citenamefont{Huang and Liu}(2019)}]{huang_moire_2019}
\bibinfo{author}{\bibfnamefont{B.}~\bibnamefont{Huang}} \bibnamefont{and}
  \bibinfo{author}{\bibfnamefont{W.~V.} \bibnamefont{Liu}},
  \bibinfo{journal}{Physical Review B} \textbf{\bibinfo{volume}{100}},
  \bibinfo{pages}{144202} (\bibinfo{year}{2019}), \bibinfo{note}{publisher:
  American Physical Society},
  \urlprefix\url{https://link.aps.org/doi/10.1103/PhysRevB.100.144202}.

\bibitem[{\citenamefont{Rossignolo and
  Dell'Anna}(2019)}]{rossignolo_localization_2019}
\bibinfo{author}{\bibfnamefont{M.}~\bibnamefont{Rossignolo}} \bibnamefont{and}
  \bibinfo{author}{\bibfnamefont{L.}~\bibnamefont{Dell'Anna}},
  \bibinfo{journal}{Physical Review B} \textbf{\bibinfo{volume}{99}},
  \bibinfo{pages}{054211} (\bibinfo{year}{2019}), \bibinfo{note}{publisher:
  American Physical Society},
  \urlprefix\url{https://link.aps.org/doi/10.1103/PhysRevB.99.054211}.

\bibitem[{\citenamefont{Roati et~al.}(2008)\citenamefont{Roati, D'Errico,
  Fallani, Fattori, Fort, Zaccanti, Modugno, Modugno, and
  Inguscio}}]{roati_anderson_2008}
\bibinfo{author}{\bibfnamefont{G.}~\bibnamefont{Roati}},
  \bibinfo{author}{\bibfnamefont{C.}~\bibnamefont{D'Errico}},
  \bibinfo{author}{\bibfnamefont{L.}~\bibnamefont{Fallani}},
  \bibinfo{author}{\bibfnamefont{M.}~\bibnamefont{Fattori}},
  \bibinfo{author}{\bibfnamefont{C.}~\bibnamefont{Fort}},
  \bibinfo{author}{\bibfnamefont{M.}~\bibnamefont{Zaccanti}},
  \bibinfo{author}{\bibfnamefont{G.}~\bibnamefont{Modugno}},
  \bibinfo{author}{\bibfnamefont{M.}~\bibnamefont{Modugno}}, \bibnamefont{and}
  \bibinfo{author}{\bibfnamefont{M.}~\bibnamefont{Inguscio}},
  \bibinfo{journal}{Nature} \textbf{\bibinfo{volume}{453}},
  \bibinfo{pages}{895} (\bibinfo{year}{2008}), ISSN \bibinfo{issn}{14764687},
  \bibinfo{note}{iSBN: 0028-0836 \_eprint: 0804.2609}.

\bibitem[{\citenamefont{Schreiber et~al.}(2015)\citenamefont{Schreiber,
  Hodgman, Bordia, Lüschen, Fischer, Vosk, Altman, Schneider, and
  Bloch}}]{schreiber_observation_2015}
\bibinfo{author}{\bibfnamefont{M.}~\bibnamefont{Schreiber}},
  \bibinfo{author}{\bibfnamefont{S.~S.} \bibnamefont{Hodgman}},
  \bibinfo{author}{\bibfnamefont{P.}~\bibnamefont{Bordia}},
  \bibinfo{author}{\bibfnamefont{H.~P.} \bibnamefont{Lüschen}},
  \bibinfo{author}{\bibfnamefont{M.~H.} \bibnamefont{Fischer}},
  \bibinfo{author}{\bibfnamefont{R.}~\bibnamefont{Vosk}},
  \bibinfo{author}{\bibfnamefont{E.}~\bibnamefont{Altman}},
  \bibinfo{author}{\bibfnamefont{U.}~\bibnamefont{Schneider}},
  \bibnamefont{and} \bibinfo{author}{\bibfnamefont{I.}~\bibnamefont{Bloch}},
  \bibinfo{journal}{Science} \textbf{\bibinfo{volume}{349}},
  \bibinfo{pages}{842} (\bibinfo{year}{2015}), ISSN \bibinfo{issn}{10959203},
  \bibinfo{note}{publisher: American Association for the Advancement of Science
  \_eprint: 1501.05661},
  \urlprefix\url{https://science.sciencemag.org/content/349/6250/842}.

\bibitem[{\citenamefont{Lüschen et~al.}(2018)\citenamefont{Lüschen, Scherg,
  Kohlert, Schreiber, Bordia, Li, Das~Sarma, and
  Bloch}}]{luschen_single-particle_2018}
\bibinfo{author}{\bibfnamefont{H.~P.} \bibnamefont{Lüschen}},
  \bibinfo{author}{\bibfnamefont{S.}~\bibnamefont{Scherg}},
  \bibinfo{author}{\bibfnamefont{T.}~\bibnamefont{Kohlert}},
  \bibinfo{author}{\bibfnamefont{M.}~\bibnamefont{Schreiber}},
  \bibinfo{author}{\bibfnamefont{P.}~\bibnamefont{Bordia}},
  \bibinfo{author}{\bibfnamefont{X.}~\bibnamefont{Li}},
  \bibinfo{author}{\bibfnamefont{S.}~\bibnamefont{Das~Sarma}},
  \bibnamefont{and} \bibinfo{author}{\bibfnamefont{I.}~\bibnamefont{Bloch}},
  \bibinfo{journal}{Physical Review Letters}  (\bibinfo{year}{2018}), ISSN
  \bibinfo{issn}{10797114}.

\bibitem[{\citenamefont{Yao et~al.}(2019)\citenamefont{Yao, Khoudli, Bresque,
  and Sanchez-Palencia}}]{yao_critical_2019}
\bibinfo{author}{\bibfnamefont{H.}~\bibnamefont{Yao}},
  \bibinfo{author}{\bibfnamefont{H.}~\bibnamefont{Khoudli}},
  \bibinfo{author}{\bibfnamefont{L.}~\bibnamefont{Bresque}}, \bibnamefont{and}
  \bibinfo{author}{\bibfnamefont{L.}~\bibnamefont{Sanchez-Palencia}},
  \bibinfo{journal}{Phys. Rev. Lett.} \textbf{\bibinfo{volume}{123}},
  \bibinfo{pages}{070405} (\bibinfo{year}{2019}), \bibinfo{note}{publisher:
  American Physical Society},
  \urlprefix\url{https://link.aps.org/doi/10.1103/PhysRevLett.123.070405}.

\bibitem[{\citenamefont{Yao et~al.}(2020)\citenamefont{Yao, Giamarchi, and
  Sanchez-Palencia}}]{yao_lieb-liniger_2020}
\bibinfo{author}{\bibfnamefont{H.}~\bibnamefont{Yao}},
  \bibinfo{author}{\bibfnamefont{T.}~\bibnamefont{Giamarchi}},
  \bibnamefont{and}
  \bibinfo{author}{\bibfnamefont{L.}~\bibnamefont{Sanchez-Palencia}},
  \bibinfo{journal}{Phys. Rev. Lett.} \textbf{\bibinfo{volume}{125}},
  \bibinfo{pages}{060401} (\bibinfo{year}{2020}), \bibinfo{note}{publisher:
  American Physical Society},
  \urlprefix\url{https://link.aps.org/doi/10.1103/PhysRevLett.125.060401}.

\bibitem[{\citenamefont{Gautier et~al.}(2021)\citenamefont{Gautier, Yao, and
  Sanchez-Palencia}}]{gautier_strongly_2021}
\bibinfo{author}{\bibfnamefont{R.}~\bibnamefont{Gautier}},
  \bibinfo{author}{\bibfnamefont{H.}~\bibnamefont{Yao}}, \bibnamefont{and}
  \bibinfo{author}{\bibfnamefont{L.}~\bibnamefont{Sanchez-Palencia}},
  \bibinfo{journal}{Phys. Rev. Lett.} \textbf{\bibinfo{volume}{126}},
  \bibinfo{pages}{110401} (\bibinfo{year}{2021}), \bibinfo{note}{publisher:
  American Physical Society},
  \urlprefix\url{https://link.aps.org/doi/10.1103/PhysRevLett.126.110401}.

\bibitem[{\citenamefont{Boers et~al.}(2007)\citenamefont{Boers, Goedeke,
  Hinrichs, and Holthaus}}]{boers_mobility_2007}
\bibinfo{author}{\bibfnamefont{D.~J.} \bibnamefont{Boers}},
  \bibinfo{author}{\bibfnamefont{B.}~\bibnamefont{Goedeke}},
  \bibinfo{author}{\bibfnamefont{D.}~\bibnamefont{Hinrichs}}, \bibnamefont{and}
  \bibinfo{author}{\bibfnamefont{M.}~\bibnamefont{Holthaus}},
  \bibinfo{journal}{Phys. Rev. A} \textbf{\bibinfo{volume}{75}},
  \bibinfo{pages}{63404} (\bibinfo{year}{2007}), \bibinfo{note}{publisher:
  American Physical Society},
  \urlprefix\url{https://link.aps.org/doi/10.1103/PhysRevA.75.063404}.

\bibitem[{\citenamefont{Modugno}(2009)}]{modugno_exponential_2009}
\bibinfo{author}{\bibfnamefont{M.}~\bibnamefont{Modugno}},
  \bibinfo{journal}{New Journal of Physics} \textbf{\bibinfo{volume}{11}},
  \bibinfo{pages}{33023} (\bibinfo{year}{2009}), \bibinfo{note}{publisher:
  \{IOP\} Publishing},
  \urlprefix\url{https://doi.org/10.1088/1367-2630/11/3/033023}.

\bibitem[{\citenamefont{An et~al.}(2021)\citenamefont{An, Padavi\'{c}, Meier,
  Hegde, Ganeshan, Pixley, Vishveshwara, and Gadway}}]{an_interactions_2021}
\bibinfo{author}{\bibfnamefont{F.~A.} \bibnamefont{An}},
  \bibinfo{author}{\bibfnamefont{K.}~\bibnamefont{Padavi\'{c}}},
  \bibinfo{author}{\bibfnamefont{E.~J.} \bibnamefont{Meier}},
  \bibinfo{author}{\bibfnamefont{S.}~\bibnamefont{Hegde}},
  \bibinfo{author}{\bibfnamefont{S.}~\bibnamefont{Ganeshan}},
  \bibinfo{author}{\bibfnamefont{J.~H.} \bibnamefont{Pixley}},
  \bibinfo{author}{\bibfnamefont{S.}~\bibnamefont{Vishveshwara}},
  \bibnamefont{and} \bibinfo{author}{\bibfnamefont{B.}~\bibnamefont{Gadway}},
  \bibinfo{journal}{Phys. Rev. Lett.} \textbf{\bibinfo{volume}{126}},
  \bibinfo{pages}{040603} (\bibinfo{year}{2021}), \bibinfo{note}{publisher:
  American Physical Society},
  \urlprefix\url{https://link.aps.org/doi/10.1103/PhysRevLett.126.040603}.

\bibitem[{\citenamefont{Kohlert et~al.}(2019)\citenamefont{Kohlert, Scherg, Li,
  Lüschen, Das~Sarma, Bloch, and Aidelsburger}}]{kohlert_observation_2019}
\bibinfo{author}{\bibfnamefont{T.}~\bibnamefont{Kohlert}},
  \bibinfo{author}{\bibfnamefont{S.}~\bibnamefont{Scherg}},
  \bibinfo{author}{\bibfnamefont{X.}~\bibnamefont{Li}},
  \bibinfo{author}{\bibfnamefont{H.~P.} \bibnamefont{Lüschen}},
  \bibinfo{author}{\bibfnamefont{S.}~\bibnamefont{Das~Sarma}},
  \bibinfo{author}{\bibfnamefont{I.}~\bibnamefont{Bloch}}, \bibnamefont{and}
  \bibinfo{author}{\bibfnamefont{M.}~\bibnamefont{Aidelsburger}},
  \bibinfo{journal}{Phys. Rev. Lett.} \textbf{\bibinfo{volume}{122}},
  \bibinfo{pages}{170403} (\bibinfo{year}{2019}), \bibinfo{note}{publisher:
  American Physical Society},
  \urlprefix\url{https://link.aps.org/doi/10.1103/PhysRevLett.122.170403}.

\bibitem[{\citenamefont{Lahini et~al.}(2009)\citenamefont{Lahini, Pugatch,
  Pozzi, Sorel, Morandotti, Davidson, and
  Silberberg}}]{lahini_observation_2009}
\bibinfo{author}{\bibfnamefont{Y.}~\bibnamefont{Lahini}},
  \bibinfo{author}{\bibfnamefont{R.}~\bibnamefont{Pugatch}},
  \bibinfo{author}{\bibfnamefont{F.}~\bibnamefont{Pozzi}},
  \bibinfo{author}{\bibfnamefont{M.}~\bibnamefont{Sorel}},
  \bibinfo{author}{\bibfnamefont{R.}~\bibnamefont{Morandotti}},
  \bibinfo{author}{\bibfnamefont{N.}~\bibnamefont{Davidson}}, \bibnamefont{and}
  \bibinfo{author}{\bibfnamefont{Y.}~\bibnamefont{Silberberg}},
  \bibinfo{journal}{Physical Review Letters}  (\bibinfo{year}{2009}), ISSN
  \bibinfo{issn}{00319007}.

\bibitem[{\citenamefont{Wang et~al.}(2020{\natexlab{b}})\citenamefont{Wang,
  Zheng, Chen, Huang, Kartashov, Torner, Konotop, and
  Ye}}]{wang_localization_2020}
\bibinfo{author}{\bibfnamefont{P.}~\bibnamefont{Wang}},
  \bibinfo{author}{\bibfnamefont{Y.}~\bibnamefont{Zheng}},
  \bibinfo{author}{\bibfnamefont{X.}~\bibnamefont{Chen}},
  \bibinfo{author}{\bibfnamefont{C.}~\bibnamefont{Huang}},
  \bibinfo{author}{\bibfnamefont{Y.~V.} \bibnamefont{Kartashov}},
  \bibinfo{author}{\bibfnamefont{L.}~\bibnamefont{Torner}},
  \bibinfo{author}{\bibfnamefont{V.~V.} \bibnamefont{Konotop}},
  \bibnamefont{and} \bibinfo{author}{\bibfnamefont{F.}~\bibnamefont{Ye}},
  \bibinfo{journal}{Nature}  (\bibinfo{year}{2020}{\natexlab{b}}), ISSN
  \bibinfo{issn}{14764687}, \bibinfo{note}{\_eprint: 2009.08131}.

\bibitem[{\citenamefont{Kraus et~al.}(2012)\citenamefont{Kraus, Lahini, Ringel,
  Verbin, and Zilberberg}}]{kraus_topological_2012}
\bibinfo{author}{\bibfnamefont{Y.~E.} \bibnamefont{Kraus}},
  \bibinfo{author}{\bibfnamefont{Y.}~\bibnamefont{Lahini}},
  \bibinfo{author}{\bibfnamefont{Z.}~\bibnamefont{Ringel}},
  \bibinfo{author}{\bibfnamefont{M.}~\bibnamefont{Verbin}}, \bibnamefont{and}
  \bibinfo{author}{\bibfnamefont{O.}~\bibnamefont{Zilberberg}},
  \bibinfo{journal}{Physical Review Letters}  (\bibinfo{year}{2012}), ISSN
  \bibinfo{issn}{00319007}, \bibinfo{note}{\_eprint: 1109.5983}.

\bibitem[{\citenamefont{Verbin et~al.}(2013)\citenamefont{Verbin, Zilberberg,
  Kraus, Lahini, and Silberberg}}]{verbin_observation_2013}
\bibinfo{author}{\bibfnamefont{M.}~\bibnamefont{Verbin}},
  \bibinfo{author}{\bibfnamefont{O.}~\bibnamefont{Zilberberg}},
  \bibinfo{author}{\bibfnamefont{Y.~E.} \bibnamefont{Kraus}},
  \bibinfo{author}{\bibfnamefont{Y.}~\bibnamefont{Lahini}}, \bibnamefont{and}
  \bibinfo{author}{\bibfnamefont{Y.}~\bibnamefont{Silberberg}},
  \bibinfo{journal}{Physical Review Letters}  (\bibinfo{year}{2013}), ISSN
  \bibinfo{issn}{00319007}, \bibinfo{note}{\_eprint: 1211.4476}.

\bibitem[{\citenamefont{Verbin et~al.}(2015)\citenamefont{Verbin, Zilberberg,
  Lahini, Kraus, and Silberberg}}]{verbin_topological_2015}
\bibinfo{author}{\bibfnamefont{M.}~\bibnamefont{Verbin}},
  \bibinfo{author}{\bibfnamefont{O.}~\bibnamefont{Zilberberg}},
  \bibinfo{author}{\bibfnamefont{Y.}~\bibnamefont{Lahini}},
  \bibinfo{author}{\bibfnamefont{Y.~E.} \bibnamefont{Kraus}}, \bibnamefont{and}
  \bibinfo{author}{\bibfnamefont{Y.}~\bibnamefont{Silberberg}},
  \bibinfo{journal}{Phys. Rev. B} \textbf{\bibinfo{volume}{91}},
  \bibinfo{pages}{64201} (\bibinfo{year}{2015}), \bibinfo{note}{publisher:
  American Physical Society},
  \urlprefix\url{https://link.aps.org/doi/10.1103/PhysRevB.91.064201}.

\bibitem[{\citenamefont{Sinelnik et~al.}(2020)\citenamefont{Sinelnik, Shishkin,
  Yu, Samusev, Belov, Limonov, Ginzburg, and
  Rybin}}]{sinelnik_experimental_2020}
\bibinfo{author}{\bibfnamefont{A.~D.} \bibnamefont{Sinelnik}},
  \bibinfo{author}{\bibfnamefont{I.~I.} \bibnamefont{Shishkin}},
  \bibinfo{author}{\bibfnamefont{X.}~\bibnamefont{Yu}},
  \bibinfo{author}{\bibfnamefont{K.~B.} \bibnamefont{Samusev}},
  \bibinfo{author}{\bibfnamefont{P.~A.} \bibnamefont{Belov}},
  \bibinfo{author}{\bibfnamefont{M.~F.} \bibnamefont{Limonov}},
  \bibinfo{author}{\bibfnamefont{P.}~\bibnamefont{Ginzburg}}, \bibnamefont{and}
  \bibinfo{author}{\bibfnamefont{M.~V.} \bibnamefont{Rybin}},
  \bibinfo{journal}{Advanced Optical Materials} \textbf{\bibinfo{volume}{8}},
  \bibinfo{pages}{2001170} (\bibinfo{year}{2020}), \bibinfo{note}{\_eprint:
  https://onlinelibrary.wiley.com/doi/pdf/10.1002/adom.202001170},
  \urlprefix\url{https://onlinelibrary.wiley.com/doi/abs/10.1002/adom.202001170}.

\bibitem[{\citenamefont{Apigo et~al.}(2019)\citenamefont{Apigo, Cheng,
  Dobiszewski, Prodan, and Prodan}}]{apigo_observation_2019}
\bibinfo{author}{\bibfnamefont{D.~J.} \bibnamefont{Apigo}},
  \bibinfo{author}{\bibfnamefont{W.}~\bibnamefont{Cheng}},
  \bibinfo{author}{\bibfnamefont{K.~F.} \bibnamefont{Dobiszewski}},
  \bibinfo{author}{\bibfnamefont{E.}~\bibnamefont{Prodan}}, \bibnamefont{and}
  \bibinfo{author}{\bibfnamefont{C.}~\bibnamefont{Prodan}},
  \bibinfo{journal}{Phys. Rev. Lett.} \textbf{\bibinfo{volume}{122}},
  \bibinfo{pages}{095501} (\bibinfo{year}{2019}), \bibinfo{note}{publisher:
  American Physical Society},
  \urlprefix\url{https://link.aps.org/doi/10.1103/PhysRevLett.122.095501}.

\bibitem[{\citenamefont{Ni et~al.}(2019)\citenamefont{Ni, Chen, Weiner, Apigo,
  Prodan, Al\`{u}, Prodan, and Khanikaev}}]{ni_observation_2019}
\bibinfo{author}{\bibfnamefont{X.}~\bibnamefont{Ni}},
  \bibinfo{author}{\bibfnamefont{K.}~\bibnamefont{Chen}},
  \bibinfo{author}{\bibfnamefont{M.}~\bibnamefont{Weiner}},
  \bibinfo{author}{\bibfnamefont{D.~J.} \bibnamefont{Apigo}},
  \bibinfo{author}{\bibfnamefont{C.}~\bibnamefont{Prodan}},
  \bibinfo{author}{\bibfnamefont{A.}~\bibnamefont{Al\`{u}}},
  \bibinfo{author}{\bibfnamefont{E.}~\bibnamefont{Prodan}}, \bibnamefont{and}
  \bibinfo{author}{\bibfnamefont{A.~B.} \bibnamefont{Khanikaev}},
  \bibinfo{journal}{Communications Physics} \textbf{\bibinfo{volume}{2}},
  \bibinfo{pages}{55} (\bibinfo{year}{2019}), ISSN \bibinfo{issn}{2399-3650},
  \urlprefix\url{https://doi.org/10.1038/s42005-019-0151-7}.

\bibitem[{\citenamefont{Cheng et~al.}(2020)\citenamefont{Cheng, Prodan, and
  Prodan}}]{cheng_experimental_2020}
\bibinfo{author}{\bibfnamefont{W.}~\bibnamefont{Cheng}},
  \bibinfo{author}{\bibfnamefont{E.}~\bibnamefont{Prodan}}, \bibnamefont{and}
  \bibinfo{author}{\bibfnamefont{C.}~\bibnamefont{Prodan}},
  \bibinfo{journal}{Phys. Rev. Lett.} \textbf{\bibinfo{volume}{125}},
  \bibinfo{pages}{224301} (\bibinfo{year}{2020}), \bibinfo{note}{publisher:
  American Physical Society},
  \urlprefix\url{https://link.aps.org/doi/10.1103/PhysRevLett.125.224301}.

\bibitem[{\citenamefont{Xia et~al.}(2020)\citenamefont{Xia, Erturk, and
  Ruzzene}}]{xia_topological_2020}
\bibinfo{author}{\bibfnamefont{Y.}~\bibnamefont{Xia}},
  \bibinfo{author}{\bibfnamefont{A.}~\bibnamefont{Erturk}}, \bibnamefont{and}
  \bibinfo{author}{\bibfnamefont{M.}~\bibnamefont{Ruzzene}},
  \bibinfo{journal}{Phys. Rev. Applied} \textbf{\bibinfo{volume}{13}},
  \bibinfo{pages}{014023} (\bibinfo{year}{2020}), \bibinfo{note}{publisher:
  American Physical Society},
  \urlprefix\url{https://link.aps.org/doi/10.1103/PhysRevApplied.13.014023}.

\bibitem[{\citenamefont{Chen et~al.}(2021)\citenamefont{Chen, Zhu, Tan, Wang,
  and Ma}}]{chen_acoustic_2021}
\bibinfo{author}{\bibfnamefont{Z.-G.} \bibnamefont{Chen}},
  \bibinfo{author}{\bibfnamefont{W.}~\bibnamefont{Zhu}},
  \bibinfo{author}{\bibfnamefont{Y.}~\bibnamefont{Tan}},
  \bibinfo{author}{\bibfnamefont{L.}~\bibnamefont{Wang}}, \bibnamefont{and}
  \bibinfo{author}{\bibfnamefont{G.}~\bibnamefont{Ma}}, \bibinfo{journal}{Phys.
  Rev. X} \textbf{\bibinfo{volume}{11}}, \bibinfo{pages}{011016}
  (\bibinfo{year}{2021}), \bibinfo{note}{publisher: American Physical Society},
  \urlprefix\url{https://link.aps.org/doi/10.1103/PhysRevX.11.011016}.

\bibitem[{\citenamefont{Gei et~al.}(2020)\citenamefont{Gei, Chen, Bosi, and
  Morini}}]{gei_phononic_2020}
\bibinfo{author}{\bibfnamefont{M.}~\bibnamefont{Gei}},
  \bibinfo{author}{\bibfnamefont{Z.}~\bibnamefont{Chen}},
  \bibinfo{author}{\bibfnamefont{F.}~\bibnamefont{Bosi}}, \bibnamefont{and}
  \bibinfo{author}{\bibfnamefont{L.}~\bibnamefont{Morini}},
  \bibinfo{journal}{Applied Physics Letters} \textbf{\bibinfo{volume}{116}},
  \bibinfo{pages}{241903} (\bibinfo{year}{2020}), \bibinfo{note}{\_eprint:
  https://doi.org/10.1063/5.0013528},
  \urlprefix\url{https://doi.org/10.1063/5.0013528}.

\bibitem[{\citenamefont{Trambly~de Laissardière
  et~al.}(2010)\citenamefont{Trambly~de Laissardière, Mayou, and
  Magaud}}]{trambly_de_laissardiere_localization_2010}
\bibinfo{author}{\bibfnamefont{G.}~\bibnamefont{Trambly~de Laissardière}},
  \bibinfo{author}{\bibfnamefont{D.}~\bibnamefont{Mayou}}, \bibnamefont{and}
  \bibinfo{author}{\bibfnamefont{L.}~\bibnamefont{Magaud}},
  \bibinfo{journal}{Nano Letters} \textbf{\bibinfo{volume}{10}},
  \bibinfo{pages}{804} (\bibinfo{year}{2010}), ISSN \bibinfo{issn}{1530-6984},
  \bibinfo{note}{publisher: American Chemical Society},
  \urlprefix\url{https://doi.org/10.1021/nl902948m}.

\bibitem[{\citenamefont{Sboychakov et~al.}(2022)\citenamefont{Sboychakov,
  Rozhkov, and Rakhmanov}}]{sboychakov_charge_2022}
\bibinfo{author}{\bibfnamefont{A.~O.} \bibnamefont{Sboychakov}},
  \bibinfo{author}{\bibfnamefont{A.~V.} \bibnamefont{Rozhkov}},
  \bibnamefont{and} \bibinfo{author}{\bibfnamefont{A.~L.}
  \bibnamefont{Rakhmanov}}, \bibinfo{journal}{JETP Letters}
  \textbf{\bibinfo{volume}{116}}, \bibinfo{pages}{729} (\bibinfo{year}{2022}),
  ISSN \bibinfo{issn}{1090-6487},
  \urlprefix\url{https://doi.org/10.1134/S0021364022602317}.

\bibitem[{\citenamefont{Breiø and Andersen}(2023)}]{breio_chern_2023}
\bibinfo{author}{\bibfnamefont{C.~N.} \bibnamefont{Breiø}} \bibnamefont{and}
  \bibinfo{author}{\bibfnamefont{B.~M.} \bibnamefont{Andersen}},
  \bibinfo{journal}{Physical Review B} \textbf{\bibinfo{volume}{107}},
  \bibinfo{pages}{165114} (\bibinfo{year}{2023}), \bibinfo{note}{publisher:
  American Physical Society},
  \urlprefix\url{https://link.aps.org/doi/10.1103/PhysRevB.107.165114}.

\bibitem[{\citenamefont{Faulstich et~al.}(2023)\citenamefont{Faulstich, Stubbs,
  Zhu, Soejima, Dilip, Zhai, Kim, Zaletel, Chan, and
  Lin}}]{faulstich_interacting_2023}
\bibinfo{author}{\bibfnamefont{F.~M.} \bibnamefont{Faulstich}},
  \bibinfo{author}{\bibfnamefont{K.~D.} \bibnamefont{Stubbs}},
  \bibinfo{author}{\bibfnamefont{Q.}~\bibnamefont{Zhu}},
  \bibinfo{author}{\bibfnamefont{T.}~\bibnamefont{Soejima}},
  \bibinfo{author}{\bibfnamefont{R.}~\bibnamefont{Dilip}},
  \bibinfo{author}{\bibfnamefont{H.}~\bibnamefont{Zhai}},
  \bibinfo{author}{\bibfnamefont{R.}~\bibnamefont{Kim}},
  \bibinfo{author}{\bibfnamefont{M.~P.} \bibnamefont{Zaletel}},
  \bibinfo{author}{\bibfnamefont{G.~K.-L.} \bibnamefont{Chan}},
  \bibnamefont{and} \bibinfo{author}{\bibfnamefont{L.}~\bibnamefont{Lin}},
  \bibinfo{journal}{Physical Review B} \textbf{\bibinfo{volume}{107}},
  \bibinfo{pages}{235123} (\bibinfo{year}{2023}), \bibinfo{note}{publisher:
  American Physical Society},
  \urlprefix\url{https://link.aps.org/doi/10.1103/PhysRevB.107.235123}.

\bibitem[{\citenamefont{González and Stauber}(2021)}]{gonzalez_magnetic_2021}
\bibinfo{author}{\bibfnamefont{J.}~\bibnamefont{González}} \bibnamefont{and}
  \bibinfo{author}{\bibfnamefont{T.}~\bibnamefont{Stauber}},
  \bibinfo{journal}{Physical Review B} \textbf{\bibinfo{volume}{104}},
  \bibinfo{pages}{115110} (\bibinfo{year}{2021}), \bibinfo{note}{publisher:
  American Physical Society},
  \urlprefix\url{https://link.aps.org/doi/10.1103/PhysRevB.104.115110}.

\bibitem[{\citenamefont{González and
  Stauber}(2020)}]{gonzalez_time-reversal_2020}
\bibinfo{author}{\bibfnamefont{J.}~\bibnamefont{González}} \bibnamefont{and}
  \bibinfo{author}{\bibfnamefont{T.}~\bibnamefont{Stauber}},
  \bibinfo{journal}{Physical Review B} \textbf{\bibinfo{volume}{102}},
  \bibinfo{pages}{081118} (\bibinfo{year}{2020}), \bibinfo{note}{publisher:
  American Physical Society},
  \urlprefix\url{https://link.aps.org/doi/10.1103/PhysRevB.102.081118}.

\bibitem[{\citenamefont{Vahedi et~al.}(2021)\citenamefont{Vahedi, Peters,
  Missaoui, Honecker, and Trambly~de Laissardière}}]{vahedi_magnetism_2021}
\bibinfo{author}{\bibfnamefont{J.}~\bibnamefont{Vahedi}},
  \bibinfo{author}{\bibfnamefont{R.}~\bibnamefont{Peters}},
  \bibinfo{author}{\bibfnamefont{A.}~\bibnamefont{Missaoui}},
  \bibinfo{author}{\bibfnamefont{A.}~\bibnamefont{Honecker}}, \bibnamefont{and}
  \bibinfo{author}{\bibfnamefont{G.}~\bibnamefont{Trambly~de Laissardière}},
  \bibinfo{journal}{SciPost Physics} \textbf{\bibinfo{volume}{11}},
  \bibinfo{pages}{083} (\bibinfo{year}{2021}), ISSN \bibinfo{issn}{2542-4653},
  \urlprefix\url{https://scipost.org/SciPostPhys.11.4.083}.

\bibitem[{\citenamefont{Schollwöck}(2005)}]{schollwock_density-matrix_2005}
\bibinfo{author}{\bibfnamefont{U.}~\bibnamefont{Schollwöck}},
  \bibinfo{journal}{Reviews of Modern Physics} \textbf{\bibinfo{volume}{77}},
  \bibinfo{pages}{259} (\bibinfo{year}{2005}), \bibinfo{note}{publisher:
  American Physical Society},
  \urlprefix\url{https://link.aps.org/doi/10.1103/RevModPhys.77.259}.

\bibitem[{\citenamefont{Gonçalves et~al.}(2023)\citenamefont{Gonçalves,
  Amorim, Riche, Castro, and Ribeiro}}]{goncalves_incommensurability_2023}
\bibinfo{author}{\bibfnamefont{M.}~\bibnamefont{Gonçalves}},
  \bibinfo{author}{\bibfnamefont{B.}~\bibnamefont{Amorim}},
  \bibinfo{author}{\bibfnamefont{F.}~\bibnamefont{Riche}},
  \bibinfo{author}{\bibfnamefont{E.~V.} \bibnamefont{Castro}},
  \bibnamefont{and} \bibinfo{author}{\bibfnamefont{P.}~\bibnamefont{Ribeiro}},
  \emph{\bibinfo{title}{Incommensurability enabled quasi-fractal order in {1D}
  narrow-band moir{\textbackslash}'e systems}} (\bibinfo{year}{2023}),
  \bibinfo{note}{arXiv:2305.03800 [cond-mat]},
  \urlprefix\url{http://arxiv.org/abs/2305.03800}.

\bibitem[{\citenamefont{Liu et~al.}(2015)\citenamefont{Liu, Ghosh, and
  Chong}}]{liu_localization_2015}
\bibinfo{author}{\bibfnamefont{F.}~\bibnamefont{Liu}},
  \bibinfo{author}{\bibfnamefont{S.}~\bibnamefont{Ghosh}}, \bibnamefont{and}
  \bibinfo{author}{\bibfnamefont{Y.~D.} \bibnamefont{Chong}},
  \bibinfo{journal}{Phys. Rev. B} \textbf{\bibinfo{volume}{91}},
  \bibinfo{pages}{014108} (\bibinfo{year}{2015}), \bibinfo{note}{publisher:
  American Physical Society},
  \urlprefix\url{https://link.aps.org/doi/10.1103/PhysRevB.91.014108}.

\bibitem[{\citenamefont{Co'}(2023)}]{co_introducing_2023}
\bibinfo{author}{\bibfnamefont{G.}~\bibnamefont{Co'}},
  \emph{\bibinfo{title}{Introducing the {Random} {Phase} {Approximation}
  {Theory}}} (\bibinfo{year}{2023}), \bibinfo{note}{arXiv:2303.05801 [cond-mat,
  physics:nucl-th]}, \urlprefix\url{http://arxiv.org/abs/2303.05801}.

\bibitem[{\citenamefont{Kuzemsky}(2015)}]{kuzemsky_variational_2015}
\bibinfo{author}{\bibfnamefont{A.~L.} \bibnamefont{Kuzemsky}},
  \bibinfo{journal}{International Journal of Modern Physics B}
  \textbf{\bibinfo{volume}{29}}, \bibinfo{pages}{1530010}
  (\bibinfo{year}{2015}), ISSN \bibinfo{issn}{0217-9792},
  \bibinfo{note}{publisher: World Scientific Publishing Co.},
  \urlprefix\url{https://www.worldscientific.com/doi/abs/10.1142/S0217979215300108}.

\bibitem[{\citenamefont{Bernasconi et~al.}(2012)\citenamefont{Bernasconi,
  Webster, Tomi\'{c}, and Harrison}}]{bernasconi_optical_2012}
\bibinfo{author}{\bibfnamefont{L.}~\bibnamefont{Bernasconi}},
  \bibinfo{author}{\bibfnamefont{R.}~\bibnamefont{Webster}},
  \bibinfo{author}{\bibfnamefont{S.}~\bibnamefont{Tomi\'{c}}},
  \bibnamefont{and} \bibinfo{author}{\bibfnamefont{N.~M.}
  \bibnamefont{Harrison}}, \bibinfo{journal}{Journal of Physics: Conference
  Series} \textbf{\bibinfo{volume}{367}}, \bibinfo{pages}{012001}
  (\bibinfo{year}{2012}), ISSN \bibinfo{issn}{1742-6596},
  \urlprefix\url{https://iopscience.iop.org/article/10.1088/1742-6596/367/1/012001}.

\bibitem[{\citenamefont{Dreuw and
  Head-Gordon}(2005)}]{dreuw_single-reference_2005}
\bibinfo{author}{\bibfnamefont{A.}~\bibnamefont{Dreuw}} \bibnamefont{and}
  \bibinfo{author}{\bibfnamefont{M.}~\bibnamefont{Head-Gordon}},
  \bibinfo{journal}{Chemical Reviews} \textbf{\bibinfo{volume}{105}},
  \bibinfo{pages}{4009} (\bibinfo{year}{2005}), ISSN \bibinfo{issn}{0009-2665},
  \bibinfo{note}{publisher: American Chemical Society},
  \urlprefix\url{https://doi.org/10.1021/cr0505627}.

\bibitem[{\citenamefont{Cazalilla et~al.}(2011)\citenamefont{Cazalilla, Citro,
  Giamarchi, Orignac, and Rigol}}]{cazalilla_one_2011}
\bibinfo{author}{\bibfnamefont{M.~A.} \bibnamefont{Cazalilla}},
  \bibinfo{author}{\bibfnamefont{R.}~\bibnamefont{Citro}},
  \bibinfo{author}{\bibfnamefont{T.}~\bibnamefont{Giamarchi}},
  \bibinfo{author}{\bibfnamefont{E.}~\bibnamefont{Orignac}}, \bibnamefont{and}
  \bibinfo{author}{\bibfnamefont{M.}~\bibnamefont{Rigol}},
  \bibinfo{journal}{Reviews of Modern Physics} \textbf{\bibinfo{volume}{83}},
  \bibinfo{pages}{1405} (\bibinfo{year}{2011}), \bibinfo{note}{publisher:
  American Physical Society},
  \urlprefix\url{https://link.aps.org/doi/10.1103/RevModPhys.83.1405}.

\bibitem[{\citenamefont{Shankar}(1994)}]{shankar_renormalization-group_1994}
\bibinfo{author}{\bibfnamefont{R.}~\bibnamefont{Shankar}},
  \bibinfo{journal}{Reviews of Modern Physics} \textbf{\bibinfo{volume}{66}},
  \bibinfo{pages}{129} (\bibinfo{year}{1994}), \bibinfo{note}{publisher:
  American Physical Society},
  \urlprefix\url{https://link.aps.org/doi/10.1103/RevModPhys.66.129}.

\end{thebibliography}

\end{document}